\documentclass[amsmath,12pt,amssymb,preprint,prd,aps,nofootinbib]{revtex4}
\usepackage{amsfonts} 
\usepackage{graphicx} 
\usepackage{epsfig}
\usepackage{multirow}
\usepackage{xcolor}
\usepackage{bm}

\newcommand{\diracslash}[1]{#1\llap{/\kern2pt}}

\newcommand{\be}{\begin{equation}}
\newcommand{\ee}{\end{equation}}
\newcommand{\bea}{\begin{eqnarray}}
\newcommand{\eea}{\end{eqnarray}}
\newcommand{\ba}[1]{\begin{array}{#1}}
\newcommand{\ea}{\end{array}}

\newcommand{\bt}{\begin{tabular}}
\newcommand{\et}{\end{tabular}}

\newcommand{\beas}{\begin{eqnarray*}}
\newcommand{\eeas}{\end{eqnarray*}}

\begin{document}
\title{Charmonium production in hot magnetized hyperonic matter\\
-- effects of baryonic Dirac sea and pseudoscalar-vector meson mixing} 
\author{Amruta Mishra}
\email{amruta@physics.iitd.ac.in}
\affiliation{Department of Physics, Indian Institute of Technology Delhi, Hauz Khas New Delhi 110016 India}
\author{Arvind Kumar}
\email{kumara@nitj.ac.in}
\affiliation{Department of Physics, Dr.\@ B.\@ R.\@ Ambedkar National Institute of Technology Jalandhar, Punjab 144008 India}
\author{S.P. Misra}
\email{misrasibaprasad@gmail.com}
\affiliation{Institute of Physics, Bhubaneswar -- 751005, India}
\vskip -1.5in
\begin{abstract}
We investigate the medium modifications of the masses of pseudoscalar 
open charm ($D$ and $\bar D$) mesons and the charmonium state 
($\psi(3770)$) in hot isospin asymmetric strange hadronic medium 
in the presence of an external magnetic field within a chiral effective
model. The in-medium partial decay widths of $\psi(3770)$ to charged
and neutral $D\bar D$ mesons are computed from the medium modifications 
of the masses of the initial and final state mesons. 
These are computed using two light quark pair creation models - 
(I) the $^3P_0$ model and (II) a field theoretical (FT)
model of composite hadrons with quark (and antiquark) constituents. 
The production cross-sections of $\psi(3770)$, arising from
scattering of the $D$ and $\bar D$ mesons, are computed 
from the relativistic Breit-Wigner spectral function
expressed in terms of the in-medium masses and the decay widths 
of the charmonium state. The effects of the 
magnetic field are considered due to the
Dirac sea (DS) of the baryons, the mixing of the 
pseudoscalar ($S$=0) and vector ($S$=1) meson 
(PV mixing), the Landau level contributions for the charged
hadrons. The anomalous magnetic moments (AMMs) of the baryons
are also taken into account in the present study.
For magnetized nuclear matter, at $\rho_B=\rho_0$, 
the effect of Dirac sea leads to inverse magnetic catalysis (IMC),
which is drop of the magnitude of the light quark condensates
(proportional to the scalar fields) with increase in the magnetic field,
contrary to the opposite effect of magnetic catalysis (MC) at $\rho_B=0$.
The inclusion of hyperons to the nuclear medium is observed 
to lead to magnetic catalysis.
There are observed to be significant effects 
from the DS and PV mixing on the properties of the charm mesons.
The production cross-sections of $\psi(3770)$ 
arising due to scattering of $D^+D^-(D^0\bar{D}^0)$ mesons 
in the hot magnetized strange hadronic matter 
are observed to have distinct peak positions, 
when the magnetic field is large. This is because the
production cross-sections have
contributions from the transverse as well as longitudinal 
components of $\psi(3770)$, 
which have non-degenerate masses due to PV ($\psi(3770)-\eta_c'$) 
mixing. 
These can have observable consequences
on the dilepton spectra as well as on the production of 
the charm mesons in ultra-relativistic peripheral 
heavy ion collision experiments, where the produced magnetic 
field is huge.
\end{abstract}
\maketitle
\newpage

\section{Introduction}
\label{intro}
Probing the properties of hadrons at finite density and/or temperature 
is an important and challenging topic of research as it can have 
implications in understanding the experimental observables 
from heavy-ion collision experiments as well as has relevance 
in the physics of compact objects \cite{Vachaspati1991,Durrer2013}.
It has been estimated that magnetic fields of the order of $2 m_\pi^2$ 
in relativistic heavy-ion collider (RHIC) at BNL, USA and 
of around $15 m_\pi^2$ in large hadron collider (LHC) at CERN 
can be produced in non-central heavy-ion collisions, which has initiated 
lots of studies on the in-medium properties of hadrons under strong magnetic 
field background \cite{ Kharzeev2008,Fukushima2008,Skovov2009}.
Strong magnetic fields are also expected to exist in the dense 
astrophysical objects, e.g, the magnetars  \cite{Roberto15,manisha23,Monika23} 
and  also play an important role in the physics of early universe 
\cite{Vachaspati1991,Durrer2013}.

The  presence of intense magnetic field leads to many interesting physical phenomenon such as magnetic catalysis (MC),   inverse magnetic catalysis (IMC),
chiral magnetic effect (CME), chiral magnetic wave  
\cite{Gusynin2012,Bali2012,Bali2013,Kharzeev2011,Huang2023}. 
The time evolution of magnetic field produced 
in heavy-ion collisions is however still debatable.
The strength of magnetic field may decay rapidly in vacuum, but the 
strongly interacting medium may cause a delay  in the decay and hence 
an increase in the life-time of magnetic field. As discussed 
in Ref. \cite{Huang2023}, the decay time of the magnetic field 
in the medium may increase due to conducting medium 
formed by charged particles and hence induction effects 
\cite{Gupta2004}.
The strong magnetic field produced during the initial stages 
of heavy-ion collisions can have imprints on the experimental 
observables related to hadrons composed of  heavy quarks as 
these are also produced during that stage, for example, 
the splitting of the directed flow $v_1$ of $D$ and $\bar{D}$ 
mesons \cite{Coci2018}.

The properties of light and heavy meson are studied in 
the external magnetic field at zero temperature and zero 
baryonic density 
\cite{Andreichikov2013,Machado2014,Carlomagno2022,Yoshida2016}. 
The non-perturbative methods such as relativistic mean field (RMF)
 models \cite{Lattimer2000,Lattimer2002,Sinha2013,Rabhi2011}, Nambu Jona Lasinio (NJL) \cite{DMenezes2009} and their polyakov extended versions (PNJL) \cite{Moreira2021,Wang2022}, quark meson coupling model \cite{Yue2008}, chiral hadronic model \cite{Dexheimer2012}, 
 chiral quark mean field model \cite{Manisha2022,Nisha2023},
 QCD sum rule approach, confined isospin density dependent model \cite{Chu2018} etc.,  
have been used in the literature to investigate the impact of external magnetic field on the quark matter, nuclear matter and compact star  properties.
The importance of magnetized Dirac sea in the calculations 
of the properties of magnetized matter and its role in observation of 
(inverse) magnetic catalysis effect have been explored in 
Refs. \cite{Menezes2009,haber2014,Arghya2018}.
In Ref. \cite{haber2014}, the vacuum to nuclear matter phase 
transition was studied using Walecka model as well as an
extended linear sigma model, 
considering Dirac sea of nucleons but neglecting the anomalous 
magnetic moments of the nucleons.
The phenomenon of magnetic catalysis, i.e., the enhancement 
of chiral condensates with increasing magnetic field, 
was observed in this work \cite{haber2014}. 
In Ref. \cite{Arghya2018}, the effects of Dirac sea of nucleons 
on the vacuum to nuclear matter phase transition were studied 
within the Walecka model considering finite anomalous magnetic 
moments of nucleons, and, using the weak magnetic field approximation 
for computing the nucleon self energy by summation of tadpole diagrams.
The anomalous 
magnetic moments (AMMs) of nucleons were observed to play a crucial 
role. The magnetic catalysis effect was observed at zero temperature 
and zero baryon density with as well as without
consideration of AMMs. However,
at finite temperature and baryonic density, 
the inverse magnetic catalysis
was observed in the presence of AMMs of the nucleons,
which was observed to be magnetic catalysis when the AMMs were ignored.
The impact of Dirac sea on the properties of nuclear matter using 
quantum hadrodynamic (QHD) model 
and performing calculations using covariant propagators 
\cite{Aguirre_Paoli_2016} for the nucleons with complete effect 
of magnetic field and anomalous magnetic moments of nucleons has been 
investigated in Ref. \cite{Aguirre2019}. The phenomenon of inverse 
magnetic catalysis 
was also reported in some lattice QCD calculations \cite{Bali2013},
where the critical temperature, $T_c$ was observed to decrease with increase
in the magnetic field. The effects of the Dirac sea for quark matter 
were investigated using NJL model in \cite{Menezes2009,DMenezes2009}.
The phenomenon of inverse magnetic catalysis was observed in a study
of magnetic effects on chiral phase transition 
in a modified AdS/QCD model \cite{Danning2016,Fang2016}.

The properties of heavy flavour mesons at finite density and temperature, 
have been explored extensively in the past,
both for zero as well as finite magnetic field background. 
In Refs. \cite{amarindam,amarvdmesonTprc}, the properties of pseudoscalar 
$D$ and $\bar{ D}$ mesons were studied in nuclear matter at finite 
temperatures within a chiral SU(3) model \cite{paper3}, 
generalized to SU(4) sector
to include the interactions of the charmed mesons. 
The properties of open charmed mesons and charmonium states 
have been studied in strange hadronic matter in Refs. 
\cite{amarvepja,Pathak12015}. 
The chiral SU(3) model has also been generalized to SU(5) sector 
to study the properties of bottom mesons in nuclear and 
strange hadronic matter \cite{Pathak22015,Pathak12014,Pathak22014}.
The chiral model has also been extended to study the impact of 
strong magnetic field on open charm mesons, open bottom mesons, 
charmonium and bottomonium properties in nuclear medium in Refs. 
\cite{Reddy2018,Dhale2018,Amal2018,charmdecay_mag,upsilon_mag}.
 The conjunction of chiral SU(3) model and QCD sum rule approach 
has also been used to study the properties of open and hidden 
heavy mesons at finite density and temperature for nonzero 
magnetic field \cite{Rajesh2020,Rajesh22020,MishraD2018,Parui2021}.
In the presence of finite strong magnetic field the phenomenon 
of PV mixing which accounts for the interaction of pseudoscalar 
mesons with the longitudinal component of vector mesons has been 
studied in context of heavy flavored mesons in Refs. 
\cite{charmdw_mag,open_charm_mag_AM_SPM,upslndw_mag}.
Recently, chiral SU(3) model is generalized to explore the impact of magnetized Dirac sea on the in-medium properties of different mesons 
\cite{Parui_D12023,Ankit_D12023,Parui22022,HQ_DW_DS_AM_SPM_2023,Sourodeep2023}.
Within effective SU(4) model the properties of charmed mesons  were studied in the nuclear matter at zero temperature, in the presence of strong external magnetic field including the impact of Dirac sea \cite{Sourodeep2023}.
The magnetized Dirac sea
along with finite anomalous magnetic moment of nucleons is observed to lead
to the phenomenon of inverse magnetic catalysis at finite densities for
symmetric as well as asymmetric nuclear matter in presence of strong
magnetic fields. The effects of the magnetized Dirac sea
on the spectral properties of light vector and axial-vector mesons 
in hot magnetized nuclear matter have also been studied recently
using a QCD sum rule approach in Ref. \cite{Parui_D32023}. 

In peripheral ultra-relativistic heavy ion collision 
experiments (where strong magnetic fields are created), since 
the matter produced is extremely dilute, we study the decay width 
of the charmonium state $\psi(3770)$, which is the 
lowest state which decays to $D\bar D$ in vacuum. 
In the present work, we investigate the in-medium masses 
of pseudoscalar $D$ and $\bar{D}$ mesons and charmonium state 
$\psi(3770)$ and their effect on the partial decay width of 
$\psi(3770)\rightarrow D\bar D$
in isospin asymmetric strange hadronic matter 
at finite temperatures in presence of a strong magnetic field.
The effects of the magnetized baryon Dirac sea contributions
as well as the PV mixing (the mixing of the pseudoscalar meson (S=0) 
and the longitudinal component of the vector meson (S=1)) on the masses
of the charmed mesons are taken into consideration.
The influence of magnetized Dirac sea of nucleons 
($p, n$) and hyperons ($\Lambda, \Sigma^{\pm,0}, \Xi^{-,0}$)
on the meson masses are explored, accounting for the 
finite anomalous magnetic moments (AMMs) of the baryons.
Additionally, the lowest Landau level contributions 
are considered for the charged open charm mesons.
The decay width of 
heavy quarkonium state to open heavy flavour mesons,
arising due to the mass modifications of the initial and final state mesons,
are computed using (I) the $^{3}P_0$ model \cite{yaouanc1,friman}
for the charm sector,
as well as (II) a field theoretical (FT) model of composite
hadrons with quark (and antiquark) constituents for both charm
and bottom sectors
\cite{spm781,spm782,amspmwg,amspm_upsilon,charmdw_mag,HQ_DW_DS_AM_SPM_2023}. 
In both of these models, the charmonium decay proceeds through creation of
a light quark-antiquark pair and the heavy charm quark (antiquark) combines
with the light antiquark (quark) to form $D\bar D$ in the final
state. The impact of PV mixing on the in-medium properties of open charm 
and charmonium sates are observed to be quite significant
in the present work. 
The well-separated non-degenerate masses of the longitudinal 
and transverse components of $\psi(3770)$ due to PV mixing is
observed to be at distinct peak positions in the production cross-sections
of $\psi(3770)$ due to scattering of the $D^+D^-(D^0\bar {D^0})$ mesons, 
when the magnetic field is large. 

Following is the outline of this paper: in section \ref{Sec.II} 
we give a brief 
description of the chiral effective model used to calculate the masses
of the charmed mesons in the finite temperature magnetized matter
at zero density as well as for isospin asymmetric (strange) 
hadronic matter. In section III, the procedure 
to calculate the in-medium masses of $D$, $\bar D$ and the charmonium  
$\psi(3770)$ and the medium modifications of 
the decay width of $\psi(3770)\rightarrow D\bar D$ due
to the mass modifications of these mesons is described. The computation
of the decay widths using the light quark-antiquark pair creation models -- 
(I) the $^3P_0$ model and (II) a field theoretical (FT) model of
composite hadrons with quark (and antiquark) constituents,
are briefly described in this section.
The results and discussions of the present study are given in
section \ref{Sec.IV} 
and the work is summarized in section \ref{Sec.V}.
\section{Chiral effective model}
\label{Sec.II}

In this section we briefly describe the chiral effective model used in the 
present work to study the in-medium properties of charmed mesons
at finite temperatures in the presence of an external magnetic field,
for zero density as well as in isospin asymmetric (strange) hadronic 
matter. A chiral SU(3) model incorporating the nonlinear 
realization of chiral symmetry \cite{paper3,Weinberg68,coleman,Bardeen69} 
and the broken scale invariance property of QCD 
\cite{paper3,hartree,kristof1} is genralized to SU(4) to include the
interactions of the charm mesons \cite{amarindam,amarvepja}.
The Lagrangian density of chiral SU(3) model is written as \cite{paper3}
\begin{equation}
{\cal L} = {\cal L}_{kin}+\sum_{W=X,Y,V,A,u} {\cal L}_{BW} + 
{\cal L}_{vec} + {\cal L}_{0} + {\cal L}_{scale\;break}+
{\cal L}_{SB} + {\cal L}_{mag}.
\label{genlag}
\end{equation}
In Eq.(\ref{genlag}), ${\cal L}_{kin}$ is the kinetic energy term, 
${\cal L}_{BW}$ 
is the baryon-meson interaction term in which the baryons-spin-0 meson 
interaction term generates the baryon masses.
${\cal L}_{vec}$  describes 
the dynamical mass generation of the vector mesons via couplings to the 
scalar mesons and contain additionally quartic self-interactions of the 
vector fields. ${\cal L}_{0}$ contains the meson-meson interaction terms, 
${\cal L}_{scale\;break}$ is the scale invariance breaking term
expressed in terms of logarthimic potential of a dilaton field,
${\cal L}_{SB}$ describes the explicit chiral symmetry breaking.
The last term ${\cal L}_{mag}$ describes the interaction of the baryons 
with the electromagnetic field.

In the present study of isospin asymmetric strange hadronic matter
at finite magnetic field, 
we adopt the mean field approximation, in which 
the meson fields are replaced by their expectation values.
The meson fields which have non-zero expectation values 
are the scalar ($\sigma, \zeta, \delta)$
and the vector fields, $\omega^\mu \rightarrow \omega \delta^{\mu 0}$,
$\rho^{\mu a}\rightarrow \rho \delta^{\mu 0} \delta_{a 3}$
and, $\phi^\mu \rightarrow \phi \delta^{\mu 0}$, and, the expectation
values of the other mesons are zero.
We also use the approximations 
$\bar \psi_i \psi_j \rightarrow \delta_{ij} \langle \bar \psi_i \psi_i
\rangle \equiv \delta_{ij} \rho^i_s $
and $\bar \psi_i \gamma^\mu \psi_j \rightarrow \delta_{ij} \delta^{\mu 0} 
\langle \bar \psi_i \gamma^ 0 \psi_i
\rangle \equiv \delta_{ij} \delta^{\mu 0} \rho_i $, 
where, $\rho_i^s$ and $\rho_i$ are the scalar and number density 
of baryon of species, $i$. 
In the above approximation, the terms which contribute 
to the baryon-meson interaction,
${\cal L}_{BW}$ of equation (\ref{genlag}) 
arise from the interactions due to the scalar (S)
and the vector (V) mesons, and, is given as
\begin{equation}
{\cal L}_{BS}+ {\cal L}_{BV}=\sum _i {\bar  \psi}^i
(g_{\sigma i} \sigma+g_{\zeta i} \zeta+g_{\delta i} \delta
-g_{\omega i}\gamma^0\omega -g_{\rho i}\gamma^0\rho
 -g_{\phi i}\gamma^0\phi)\psi^i. 
\label{BS_BV}
\end{equation}
The effective mass of the baryon of species $i$ is obtained from
the in-medium values of the scalar-isoscalar non-strange ($\sigma$) 
and strange ($\zeta$) mesons and the scalar-isovector ($\delta$)
meson and is given as
\begin{equation}
m_{i}^{*} = - g_{\sigma i}\sigma- g_{\zeta i}\zeta - g_{\delta i} \delta,
\label{meffi}
\end{equation}
and the effective chemical potential for $i$-th baryon, due to the 
interaction with the vector mesons ($\omega$, $\rho$ and $\phi$) 
is given as
\begin{equation}
\mu^*_i=\mu_i-g_{\omega i}\omega-g_{\rho i}\rho-g_{\phi i}\phi
\label{mueffi}
\end{equation}
The other terms in the Lagrangian density of equation (\ref{genlag}) 
are given as
\begin{align}
{\cal L}_{vec} &= \frac{1}{2} \frac{\chi^2}{\chi_0^2}\Big(
m_{\omega}^{2} \omega^ 2+m_{\rho}^{2} \rho^ 2+m_{\phi}^{2} \phi^ 2
\Big) +g_4^4 (\omega^4 +2 \phi^4+6 \omega^2 \rho^2+\rho^4)\\
{\cal L}_{0} & = - \frac{ 1 }{ 2 } k_0 \chi^2
(\sigma^2+\zeta^2+\delta^2) + k_1 (\sigma^2+\zeta^2+\delta^2)^2
\nonumber \\
     &+ k_2 ( \frac{ \sigma^4}{ 2 } + \frac{\delta^4}{2} + \zeta^4
 +3 \sigma^2 \delta^2)
     + k_3 \chi (\sigma^2 - \delta^2) \zeta - k_4 \chi^4 \\
\label{L_0}
{\cal L}_{scale \; break}&= -\frac{1}{4} \chi^{4} 
{\rm ln} \frac{\chi^{4}}{\chi_{0}^{4}} + \frac{d}{3} \chi^{4} 
{\rm ln} \Bigg( \frac{\left( \sigma^{2} - \delta^{2}\right)\zeta }
{\sigma_{0}^{2} \zeta_{0}} \Big( \frac{\chi}{\chi_{0}}\Big) ^{3}\Bigg), \\
\label{scalebreak}
{\cal L} _{SB} & =  - \left( \frac{\chi}{\chi_{0}}\right) ^{2} 
\left[ m_{\pi}^{2} 
f_{\pi} \sigma + \left( \sqrt{2} m_{K}^{2}f_{K} - \frac{1}{\sqrt{2}} 
m_{\pi}^{2} f_{\pi} \right) \zeta \right], 
\end{align}
and,
\be 
{\cal L}_{mag}=-{\bar {\psi_i}}q_i 
\gamma_\mu A^\mu \psi_i
-\frac {1}{2} \kappa_i {\bar {\psi_i}} \sigma ^{\mu \nu}F_{\mu \nu}
\psi_i
-\frac{1}{4} F^{\mu \nu} F_{\mu \nu}.
\label{lmag}
\ee
In the above, the parameters $k_0$, $k_2$ and $k_4$ are fitted
to ensure extremum in the vacuum for the equations of  motion
for scalar fields $\sigma$, $\zeta$ and the dilaton field $\chi$, 
$k_1$ is fitted to reproduce
the mass of $\sigma$ to be of the order of 500 MeV, $k_3$
is fitted from the $\eta$ and $\eta'$ masses, and the value of
the $\chi$ in vacuum is fitted so that the pressure $p=0$ 
at the nuclear matter saturation density.
In equation (\ref{lmag}), $q_i$ is the electric charge of 
the $i^{th}$ baryon described by field $\psi_i$.
The second term is a tensorial interaction term expressed
in terms of the electromagnetic field tensor, $F_{\mu \nu}$
($=\partial_\mu A_\nu -\partial_\nu A_\mu$), 
$\sigma ^{\mu \nu}=\frac{i}{2}[\gamma^\mu,\gamma^\nu]$, 
and $\kappa_i$, the anomalous magnetic moment (AMM) 
of $i$-{th} baryon.

The concept of broken scale invariance leading to the trace anomaly 
in QCD, $\theta_{\mu}^{\mu} = \frac{\beta_{QCD}}{2g} 
{G^a}_{\mu\nu} G^{\mu\nu a}$, where $G_{\mu\nu}^{a} $ is the 
gluon field strength tensor of QCD, is simulated in the effective 
Lagrangian at tree level through the introduction of 
the scale breaking term \cite{sche1,heide1} given by
equation (\ref{scalebreak}).
Equating the trace of the energy momentum tensor 
with that of the present chiral model
gives the relation of the dilaton field to the scalar gluon condensate.
%
The explicit symmetry breaking term in the chiral model 
is given by
\begin{eqnarray}
{\cal L} _{SB} & = & 
-\left( \frac{\chi}{\chi_{0}}\right) ^{2} 
{\rm Tr} \left [ {\rm diag} \left (
\frac{m_{\pi}^{2} f_{\pi}}{2} (\sigma+\delta), 
\frac{m_{\pi}^{2} f_{\pi}}{2} (\sigma-\delta), 
\Big( \sqrt{2} m_{K}^{2}f_{K} - \frac{1}{\sqrt{2}} 
m_{\pi}^{2} f_{\pi} \Big) \zeta \right) \right ],
\label{lag_SB_MFT}
\end{eqnarray}
which reduces to eqn. (8), and in QCD, 
this term is given as ${\cal L}_{SB}^{QCD}
=-Tr[{\rm diag}(m_u \bar u u,m_d \bar d d, m_s \bar s s)]$.
In the above, the matrices, whose traces give the 
corresponding Lagrangian densities
have been explicitly written down, and a comparison of their 
matrix elements relates the quark condensates 
to the values of the scalar fields 
as given by \cite{am_vecqsr}
\begin{eqnarray}
m_u\langle \bar u u \rangle 
& = &\Big (\frac{\chi}{\chi_{0}}\Big) ^{2} 
\frac{m_{\pi}^{2} f_{\pi}}{2} (\sigma+\delta),\;
m_d \langle \bar d d \rangle
=\left( \frac{\chi}{\chi_{0}}\right) ^{2} 
\frac{m_{\pi}^{2} f_{\pi}}{2} (\sigma-\delta),
\nonumber \\
m_s\langle \bar s s \rangle 
& =& \left( \frac{\chi}{\chi_{0}}\right) ^{2} 
\big( \sqrt {2} m_{K}^{2}f_{K} - \frac {1}{\sqrt {2}} 
m_{\pi}^{2} f_{\pi} \big) \zeta.
\label{qqbar}
\end{eqnarray}
%
%
The thermodynamic potential for the magnetized hadronic medium 
is written as
\begin{align}
\label{thermo_total}
\Omega = \Omega_{\text{DS}}+ \Omega_{med}
-{\cal L}_{vec} - {\cal L}_0 - {\cal L}_{scale\;break} - {\cal L}_{SB}.
\end{align}
In the equation (\ref{thermo_total}), $\Omega_{\text{DS}}$ 
and $\Omega_{med}$ denote the contributions of Dirac sea (DS) 
and Fermi sea (medium part)
of the (charged and neutral) baryons to the thermodynamic potential,
with the masses and chemical potentials of the baryons, modified
due to their interactions with the scalar and vector fields,
given by equations (\ref{meffi}) and (\ref{mueffi}) respectively. 
For spin-1/2 charged baryons ($i=p,\Sigma^{\pm},\Xi^-$), 
these are given as
\cite{haber2014,Aguirre_Paoli_2016,Aguirre2019}
\begin{align}
\Omega_{\text{DS}}^{charged} = \sum_{i} \Omega_{\text{DS}}^{i}
 =-\sum_{i}\frac{|q_iB|}{2\pi} 
\sum_{s = \pm 1}\sum_{\nu=0}^{\infty} (2-\delta_{\nu 0}) 
\int_{-\infty}^{\infty} \frac{dk_z}{2\pi} \epsilon^i_{k,\nu,s}.
\label{ther_pot_Dirac_sea_charged_baryons}
\end{align}
and
\begin{align}
\Omega_{med}^{charged} & =\sum_{i}\Omega_{med}^{i}
=-T\sum_{i}
\frac{|q_i| B}{2\pi}  
\sum_{s = \pm 1}\sum_{\nu=0}^{\infty} (2-\delta_{\nu 0}) 
\nonumber \\ &
\int_{-\infty}^\infty \frac {d k_z}{2\pi} \, \biggl\{{\rm ln}
\left( 1+e^{-\beta ( \epsilon^{i}_{k,\nu, s} - \mu^{*}_{i} )}\right) 
+ {\rm ln}\left( 1+e^{-\beta ( \epsilon^{i}_{k,\nu, s}+\mu^{*}_{i} )}
\right) \biggr\},
\label{ther_pot_Fermi_sea_charged_baryons}
\end{align}
where, where $\beta=1/T$ and $\epsilon^i_{k,\nu,s}$ 
is the single particle energy 
of the $i$-th charged baryon given as
\begin{equation}
\epsilon^i_{k,\nu,s} = \sqrt{k_z^2 + \left(\sqrt{2\nu|q_iB|
+m_i^{*2}}-s{\kappa_i} B\right)^2}.
\end{equation}

The thermodynamic potentials corresponding to the Dirac sea
and the Fermi sea of the neutral baryons ($i=n,\Lambda,\Sigma^0,\Xi^0$)
are given as
\cite{haber2014,Aguirre_Paoli_2016,Aguirre2019}
\begin{align}
\Omega_{\text{DS}}^{Neutral} =
\sum _i \Omega_{\text{DS}}^{i} =
 -\sum _i \sum_{s = \pm 1}  \int  
\frac{d^3k}{\left(2\pi\right)^3} \epsilon^i_{k,s},
\label{ther_pot_Dirac_sea_neutral_baryons}
\end{align}
and,
\begin{equation}
\Omega_{med}^{Neutral}= 
\sum_i\Omega_{med}^{i}= 
-T \sum_i \sum_{s=\pm 1} \int \frac{d^3k}{(2\pi)^3} 
\biggl\{{\rm ln}
\left( 1+e^{-\beta (\epsilon^{i}_{k,s} - \mu^{*}_{i} )}\right) \\
+ {\rm ln}\left( 1+e^{-\beta (\epsilon^{i}_{k,s}+\mu^{*}_{i})}
\right) \biggr\},
\label{ther_pot_Fermi_sea_neutral_baryons}
\end{equation}
with the single particle energy for the $i$-th neutral baryon given as
\begin{equation}
\epsilon^i_{k,s} = \sqrt{k_z^2 
+ \left ( \sqrt{k_x^2 +k_y^2+ {m_i^{*}}^2}-s\kappa_i B\right)^2}.
\end{equation}

%
For given values of the baryon density, $\rho_B=\sum_i \rho_i$,
with $\rho_i$ as the number density of the $i$-th baryon,
the isospin asymmetry parameter
$\eta = -\frac{\Sigma_i I_{3i} \rho_{i}}{\rho_{B}}$, with
$I_{3i}$ as the 3$^{rd}$ component of isospin quantum number
of the $i$-th baryon, the strangeness fraction, 
$f_s = \frac{\Sigma_i \vert s_{i} \vert \rho_{i}}{\rho_{B}}$,
with $s_i$ number of strange quarks in the $i$-th baryon,
temperature, $T$ and magnetic field, $B$, 
the minimisation of the thermodynamic potential
leads to the coupled equations for the scalar fields 
($\sigma,\zeta,\delta$), 
the dilaton field $\chi$, 
and the vector fields ($\omega$,$\rho$, $\phi$)
as given by
\begin{eqnarray}
&& k_{0}\chi^{2}\sigma-4k_{1}\left( \sigma^{2}+\zeta^{2}
+\delta^{2}\right)\sigma-2k_{2}\left( \sigma^{3}+3\sigma\delta^{2}\right)
-2k_{3}\chi\sigma\zeta \nonumber\\
&-&\frac{d}{3} \chi^{4} \bigg (\frac{2\sigma}{\sigma^{2}-\delta^{2}}\bigg )
+\left( \frac{\chi}{\chi_{0}}\right) ^{2}m_{\pi}^{2}f_{\pi}
-{\sum_i} g_{\sigma i}\rho^{i}_{s} = 0
\label{sigma}
\end{eqnarray}
\begin{eqnarray}
&& k_{0}\chi^{2}\zeta-4k_{1}\left( \sigma^{2}+\zeta^{2}+\delta^{2}\right)
\zeta-4k_{2}\zeta^{3}-k_{3}\chi\left( \sigma^{2}-\delta^{2}\right)\nonumber\\
&-&\frac{d}{3}\frac{\chi^{4}}{\zeta}+\left(\frac{\chi}{\chi_{0}} \right)
^{2}\left[ \sqrt{2}m_{K}^{2}f_{K}-\frac{1}{\sqrt{2}} m_{\pi}^{2}f_{\pi}\right]
 -{\sum_i} g_{\zeta i}\rho_{s}^{i} = 0
\label{zeta}
\end{eqnarray}
\begin{eqnarray}
& & k_{0}\chi^{2}\delta-4k_{1}\left( \sigma^{2}+\zeta^{2}+\delta^{2}\right)
\delta-2k_{2}\left( \delta^{3}+3\sigma^{2}\delta\right) +k_{3}\chi\delta
\zeta \nonumber\\
& + &  \frac{2}{3} d \left( \frac{\delta}{\sigma^{2}-\delta^{2}}\right)
-{\sum_i} g_{\delta i}\rho_{s}^{i} = 0
\label{delta}
\end{eqnarray}
\begin{eqnarray}
& & k_{0}\chi \left( \sigma^{2}+\zeta^{2}+\delta^{2}\right)-k_{3}
\left( \sigma^{2}-\delta^{2}\right)\zeta + \chi^{3}\left[1
+{\rm {ln}}\left( \frac{\chi^{4}}{\chi_{0}^{4}}\right)  \right]
+(4k_{4}-d)\chi^{3}
\nonumber\\
& - & \frac{4}{3} d \chi^{3} {\rm {ln}} \Bigg ( \bigg (\frac{\left( \sigma^{2}
-\delta^{2}\right) \zeta}{\sigma_{0}^{2}\zeta_{0}} \bigg )
\bigg (\frac{\chi}{\chi_0}\bigg)^3 \Bigg )
+\frac{2\chi}{\chi_{0}^{2}}\left[ m_{\pi}^{2}
f_{\pi}\sigma +\left(\sqrt{2}m_{K}^{2}f_{K}-\frac{1}{\sqrt{2}}
m_{\pi}^{2}f_{\pi} \right) \zeta\right]  = 0
\label{chi}
\end{eqnarray}
\begin{align}
\left (\frac{\chi}{\chi_{0}}\right) ^{2}m_{\omega}^{2}\omega
+g_{4}^{4}\left(4{\omega}^{3}+12{\rho}^2{\omega}\right)
-{\sum_i} g_{\omega i}\rho_{i}  = 0 ,
\label{omega} 
\end{align}
\begin{align}
\left (\frac{\chi}{\chi_{0}}\right) ^{2}m_{\rho}^{2}\rho
+g_{4}^{4}\left(4{\rho}^{3}+12{\omega}^2{\rho}\right)
-{\sum_i} g_{\rho i} \rho_{i}  = 0,
\label{rho}
\end{align}
\begin{align}
\left (\frac{\chi}{\chi_{0}}\right) ^{2}m_{\phi}^{2}\phi
+ 8 g_{4}^{4}{\phi}^{3}-{\sum_i} g_{\phi i}\rho_{i}  = 0 ,
\label{phi}
\end{align}
where, $\rho_i$ is the number density of the $i$-th baryon,
given by the expression
\begin{align}
\rho_{i}=\frac{|q_{i}B|}{2\pi^2} \sum_{s = \pm 1}
\sum_{\nu=0}^{\infty}
(2-\delta_{\nu0})
 \int^{\infty}_{0}
dk_{z}\Bigg[
\frac{1}{1+e^{\beta(  \epsilon^i_{k,\nu, s} 
-\mu^{*}_{i})}} 
-
\frac{1}{1+e^{\beta(  \epsilon^i_{k,\nu, s} 
+\mu^{*}_{i})}} 
\Bigg],
\label{rhovp_ch}
\end{align}
for charged baryons and
\begin{align}
\rho_{i}=  \sum_{s = \pm 1}
 \int
\frac{d^3k}{(2\pi)^3}\Bigg[
\frac{1}{1+e^{\beta(  \epsilon^i_{k, s} 
-\mu^{*}_{i})}} 
-
\frac{1}{1+e^{\beta(  \epsilon^i_{k, s} 
+\mu^{*}_{i})}} 
\Bigg],
\label{rhovp_neutral}
\end{align}
for the neutral baryons.
In the above equations, the scalar density of the $i$-th baryon, 
$\rho_s^i=\langle {\bar \psi^i} \psi^i \rangle
={\partial \Omega}/{\partial {m_i^*}}$. 
The expectation values of the scalar fields ($\sigma$, $\zeta$
and $\delta$) are obtained by solving the equations self-consistently,
since these values depend on the baryon scalar 
densities, $\rho_s^i$, which, in turn, are functions 
of the scalar field (through the effective baryon masses). 
Neglecting the contribution from the Dirac sea 
to the thermodynamic potential, 
the expression for the scalar density of the $i$-th baryon, 
$\rho_s^i$
%
is given as 
\begin{align}
\rho_s^i  \equiv  {\rho_s^i}^{med} &=\frac{|q_i| B}{2\pi^2} 
\sum_{s = \pm 1}\sum_{\nu=0}^{\infty} (2-\delta_{\nu 0}) 
m_i^* \int_{0}^\infty d k_z
\frac{\sqrt{{m_i^*}^2+2\nu |q_i| B}-s\kappa_i B}{\epsilon^{i}_{k,\nu,s}
\sqrt{{m_i^*}^2+2\nu |q_i| B}}
\nonumber \\ &
\times \Bigg [
\frac {1}{1+e^{\beta (\epsilon^{i}_{k,\nu, s}-\mu^{*}_i)}}
+\frac {1}{1+e^{\beta (\epsilon^{i}_{k,\nu, s}+\mu^{*}_i)}}
\Bigg],
\label{rho_s_i_ch_med}
\end{align}
for the charged baryons and
\begin{align}
\rho_s^i & \equiv {\rho_s^i}^{med} = 
m_i^* \int \frac{ d^3 k}{(2\pi)^3} 
\sum_{s = \pm 1}
\frac{\sqrt{k_x^2 +k_y^2+ {m_i^{*}}^2}-s\kappa_i B}
{\epsilon^i_{k,s}\sqrt{k_x^2 
+k_y^2+ {m_i^{*}}^2}}
\Bigg [
\frac {1}{1+e^{\beta (\epsilon^{i}_{k, s}-\mu^{*}_i)}}
+\frac {1}{1+e^{\beta (\epsilon^{i}_{k, s}+\mu^{*}_i)}}
\Bigg],
\label{rho_s_i_neutral_med}
\end{align}
for the neutral baryons.
Within the chiral effective model, the effects of the 
magnetic field have been studied for isospin asymmetric nuclear
matter on the open charm \cite{Reddy2018}, open bottom
\cite{Dhale2018}, charmonium \cite{Amal2018} and bottomonium 
\cite {upsilon_mag} states, considering the effects of the nucleon AMMs. 
However, these studies were without incorporating 
the effects of the nucleon Dirac sea, and, were considering 
only the contributions from the medium parts of the scalar densities
for the charged and neutral baryons given by equations
(\ref{rho_s_i_ch_med}) and (\ref{rho_s_i_neutral_med}),
while solving the values of the scalar fields ($\sigma$,
$\zeta$, $\delta$ and $\chi$), along with the vector fields,
from the equations (\ref{sigma})-(\ref{phi}).
In the present work of the study the charmed mesons 
in hot strange hadronic matter in the presence of 
an external magnetic field,
the values of the scalar fields are solved 
including the contributions to the scalar densities
from the Dirac sea of baryons. 
As may be noted, the expressions corresponding to the contributions
to the thermodynamic potential from the Dirac sea
for the charged and neutral baryons, given by equations 
(\ref{ther_pot_Dirac_sea_charged_baryons})
and (\ref{ther_pot_Dirac_sea_neutral_baryons})
are divergent and need renormalization. In Ref. \cite{haber2014},
the nuclear matter in strong magnetic fields was studied within 
the Walecka model as well as an extended linear sigma model, ignoring
the anomalous magnetic moments (AMMs) of the nucleons and the 
renormalization of the Dirac sea contribution due to the charged baryon,
i.e., proton, was carried out by using proper time method.
Due to its smallness, the pure vacuum ($\rho_B$=0, T=0) 
contribution corresponding to zero magnetic field was neglected, 
and the magnetized Dirac sea (the part with non-zero B)
was observed to lead to magnetic catalysis. There was no contribution
from the neutron due to the Dirac sea, since the AMM of the neutron
was not considered. Including the AMMs of the nucleons into account,
the covariant fermion propagator in the presence of strong magnetic
fields at finite temperature and density has been calculated  
in Ref. \cite{Aguirre_Paoli_2016}, which has been generalized
to include the effects of the Dirac sea in Ref. \cite{Aguirre2019}.
The value of the scalar field, $\sigma$ is calculated
through the minimisation of the thermodynamic potential 
\cite{haber2014}, from which the effective mass 
of the nucleon is obtained. 
There is observed to be increase in the nucleon mass
(corresponding to increase in the light quark condensates)
with the increase in the magnetic field,
when the AMMs of the nucleons are not considered \cite{haber2014}, 
an effect called the magnetic catalysis (MC). On the other hand,  
the inclusion of the AMMs is observed to lead to the opposite 
behaviour for the light quark condensates, i.e., the
effect of inverse magnetic catalysis (IMC) \cite{Arghya2018}.
In Ref. \cite{Arghya2018}, the baryon propagator
in the presence of a magnetic field was calculated 
accounting for the AMMs of the nucleons within Walecka model, 
including the effects of the Dirac sea 
by summing up the scalar ($\sigma$) and vector ($\omega$) 
tadpole diagrams in the weak field approximation. 
In the present study of 
magnetized hot strange hadronic matter
within the chiral effective model,
the contribution of the
magnetized Dirac sea to the baryon self energy
is calculated by summing the scalar ($\sigma$, $\zeta$
and $\delta$) and vector ($\omega$, $\rho$ and $\phi$)
tadpole diagrams, arising from the  baryon-meson interaction
given by equation (\ref{BS_BV})  within the weak field approximation.
The propagator for the $i$-th baryon, $S_B^i(p)$ is given by 
\cite{Nieves2004,Arghya2018}
\begin{equation}
\Bigg [\gamma^\mu p_\mu -\frac{i}{2}q_i F^{\mu \nu}\gamma_\mu 
\frac {\partial}{\partial p^\nu}-m_i^* -\frac{1}{2} \kappa_i
F^{\mu \nu}\sigma_{\mu \nu} \Bigg] S_B^i(p)=1,
\end{equation}
where, $q_i$ and $\kappa_i$ are the electric charge
and AMM of the $i$-th baryon, $m_i^*$ is its effective baryon mass, 
given by equation (\ref{meffi}).
The baryon propagator is expanded in powers of 
$q_i B$ and $\kappa_i B$, retaining upto quadratic order
in the weak field approximation. 
The divergent expression for the
self-energy for pure vacuum (B=0), $\Sigma_{DS,B=0}$ 
is neglected due to its smallness \cite{haber2014,Arghya2018}. 
The dimensional regularization is used to separate
the divergent contribution of the magnetized Dirac sea contribution,
$\Sigma_{DS,B}^{div}$, followed by ${\overline {MS}}$ prescription 
and determination of the scale parameter, $\Lambda$, using the condition 
$\Sigma_{DS,B} ^{div} (m_i^*=m_i)=0$.
In the present study of magnetized strange hadronic matter
at finite temperature, within the chiral effective model,
the contribution to the self energy
of the $i$-th baryon due to the Dirac sea is given as
\begin{equation}
\Sigma ^{DS,i} 
=-\Big (\frac{g_{\sigma i}^2}{m_\sigma^2} +
\frac{g_{\zeta i}^2}{m_\zeta^2} +
\frac{g_{\delta i}^2}{m_\delta^2}\Big)\rho_s^{DS,i}
\equiv -A_i \rho_s^{DS,i},
\label{self_energy_i_DS}
\end{equation}
where,
\begin{equation}
\rho_s^{DS,i}=-\frac{1}{4\pi^2} \Bigg [ \frac{(q_i B)^2}{3 m_i^*}
+\{(\kappa_i B)^2 m_i^* +(|q_i|B)(\kappa_i B)\} 
\Big (\frac{1}{2} +2 {\rm ln} \Big (\frac{m_i^*}{m_i}\Big)
\Big) \Bigg].
\label{rhosi_DS}
\end{equation}
The effects of the Dirac sea of the baryons
are taken into consideration by adding the Dirac sea contributions 
to the scalar densities for the baryons.
The contribution from the magnetized Dirac sea to the scalar density 
of the $i$-th baryon, with electric charge, $q_i$, 
the anomalous magnetic moment, $\kappa_i$ and effective
mass, $m_i^*$, is given by equation (\ref{rhosi_DS}). 
As has already been mentioned, the values of scalar 
fields $\sigma$, $\zeta$, $\delta$, the dilaton field 
$\chi$, along with the vector fields, are obtained by 
solutions of their coupled equations of motion as
given by equations (\ref{sigma})-(\ref{phi}), 
for given values of the temperature ($T$) and magnetic field ($B$),
for zero baryon density, as well as, for (strange) hadronic matter
for given values of baryon density ($\rho_B$), 
isospin asymmetry parameter ($\eta$) and strangeness fraction ($f_s$).
These are used to obtain the in-medium masses of the open charm 
($D$ and $\bar D$) mesons, the charmonium $\psi(3770)$,
and the partial decay width of $\psi(3770)\rightarrow D\bar D$ 
in magnetized isospin asymmetric strange hadronic matter 
at finite temperatures. The production cross-sections
of $\psi(3770)$ due to the scattering of the charged
and neutral $D\bar D$ pairs are obtained using the in-medium masses
and the decay widths in the present work.

\section{Charm mesons in hot isospin asymmetric magnetized 
strange hadronic matter}
\label{Sec.III}
\subsection{Masses of open charm and charmonium states}
In magnetized isospin asymmetric strange hadronic matter at finite 
temperatures, the medium modifications 
of the masses of the open charm ($D$ and $\bar D$) mesons arise 
due to their interactions with the nucleons, hyperons and the
scalar mesons ($\sigma$ and $\delta$), whereas, the in-medium 
mass of the charmonium state $\psi(3770)$ is calculated from
the medium change of a dilaton field (which mimics the gluon condensates
of QCD) within a chiral effective model \cite{amarvepja}. 
The interaction Lagrangian density 
for $D$ and $\bar{D}$ mesons is given as \cite{amarvepja} 
\begin{align}
\cal L _{D(\bar D)} & =  -\frac {i}{8 f_D^2} \Big [3\Big (\bar p \gamma^\mu p
+\bar n \gamma ^\mu n \Big) 
\Big(\Big({D^0} (\partial_\mu \bar D^0) - (\partial_\mu {{D^0}}) {\bar D}^0 \Big )
+\Big(D^+ (\partial_\mu D^-) - (\partial_\mu {D^+})  D^- \Big )\Big )
\nonumber \\
& +
\Big (\bar p \gamma^\mu p -\bar n \gamma ^\mu n \Big) 
\Big( \Big({D^0} (\partial_\mu \bar D^0) - (\partial_\mu {{D^0}}) {\bar D}^0 \Big ) 
- \Big( D^+ (\partial_\mu D^-) - (\partial_\mu {D^+})  D^- \Big )\Big )\nonumber\\
&+ 2\Big(\bar{\Lambda}^{0}\gamma^{\mu}\Lambda^{0}\Big)
 \Big( \Big({D^0} (\partial_\mu \bar D^0)
 -(\partial_\mu {{D^0}}) {\bar D}^0 \Big)
+ \Big(D^+ (\partial_\mu D^-) - (\partial_\mu D^+)  D^- \Big) \Big)\nonumber\\
 &+ 2 \Big(\Big(\bar{\Sigma}^{+}\gamma^{\mu}\Sigma^{+}
 + \bar{\Sigma}^{-}\gamma^{\mu}\Sigma^{-}\Big)
 \Big(\Big({D^0} (\partial_\mu \bar D^0)
 -(\partial_\mu {{D^0}}) {\bar D}^0 \Big)
+ \Big(D^+ (\partial_\mu D^-) - (\partial_\mu D^+)  D^- \Big)\Big)\nonumber\\
&+ \Big(\bar{\Sigma}^{+}\gamma^{\mu}\Sigma^{+}
 - \bar{\Sigma}^{-}\gamma^{\mu}\Sigma^{-}\Big)
 \Big(\Big({D^0} (\partial_\mu \bar D^0)
 -(\partial_\mu {{D^0}}) {\bar D}^0 \Big)
- \Big(D^+ (\partial_\mu D^-) - (\partial_\mu D^+)  D^- \Big)\Big)\Big)\nonumber\\
&+2\Big(\bar{\Sigma}^{0}\gamma^{\mu}\Sigma^{0}\Big)
 \Big(\Big({D^0} (\partial_\mu \bar D^0)
 -(\partial_\mu {{D^0}}) {\bar D}^0 \Big)
+ \Big(D^+ (\partial_\mu D^-) - (\partial_\mu D^+)  D^- \Big) \Big)\nonumber\\
 &+ \Big(\bar{\Xi}^{0}\gamma^{\mu}\Xi^{0}
 + \bar{\Xi}^{-}\gamma^{\mu}\Xi^{-}\Big)
 \Big(\Big({D^0} (\partial_\mu \bar D^0)
 -(\partial_\mu {{D^0}}) {\bar D}^0 \Big)
+ \Big(D^+ (\partial_\mu D^-) - (\partial_\mu D^+)  D^- \Big)\Big)\nonumber\\
&+\Big(\bar{\Xi}^{0}\gamma^{\mu}\Xi^{0}
 - \bar{\Xi}^{-}\gamma^{\mu}\Xi^{-}\Big)
 \Big(\Big({D^0} (\partial_\mu \bar D^0)
 -(\partial_\mu {{D^0}}) {\bar D}^0 \Big)
- \Big(D^+ (\partial_\mu D^-) - (\partial_\mu D^+)  D^- \Big)\Big)\Big ]\nonumber\\
&+ \frac{m_D^2}{2f_D} \Big [ 
(\sigma +\sqrt 2 \zeta_c)\big (\bar D^0 { D^0}+(D^- D^+) \big )
 +\delta (\big (\bar D^0 { D^0})-(D^- D^+) \big )
\Big ] \nonumber \\
& -  \frac {1}{f_D}\Big [ 
(\sigma +\sqrt 2 \zeta_c )
\Big ((\partial _\mu {{\bar D}^0})(\partial ^\mu {D^0})
+(\partial _\mu {D^-})(\partial ^\mu {D^+}) \Big )
  +  \delta
\Big ((\partial _\mu {{\bar D}^0})(\partial ^\mu {D^0})
-(\partial _\mu {D^-})(\partial ^\mu {D^+}) \Big )
\Big ]
\nonumber \\
&+  \frac {d_1}{2 f_D^2}(\bar p p +\bar n n +\bar{\Lambda}^{0}\Lambda^{0}+\bar{\Sigma}^{+}\Sigma^{+}+\bar{\Sigma}^{0}\Sigma^{0}
+\bar{\Sigma}^{-}\Sigma^{-}+\bar{\Xi}^{0}\Xi^{0}+\bar{\Xi}^{-}\Xi^{-})
 \big ( (\partial _\mu {D^-})
 (\partial ^\mu {D^+})
 \nonumber\\
&+(\partial _\mu {{\bar D}^0})(\partial ^\mu {D^0})
\big )
+ \frac {d_2}{2 f_D^2} \Big [
\Big(\bar p p+\frac{1}{6}\bar{\Lambda}^{0}\Lambda^{0}+\bar{\Sigma}^{+}\Sigma^{+}+\frac{1}{2}\bar{\Sigma}^{0}\Sigma^{0}\Big)
(\partial_\mu {\bar D}^0)(\partial^\mu {D^0})\nonumber\\
&+\Big(\bar n n+\frac{1}{6}\bar{\Lambda}^{0}\Lambda^{0}+\bar{\Sigma}^{-}\Sigma^{-}+\frac{1}{2}\bar{\Sigma}^{0}\Sigma^{0}\Big) (\partial_\mu D^-)(\partial^\mu D^+)\Big ],
\label{ldb}
\end{align}
where, the first term is the vectorial Weinberg Tomozawa 
interaction term,
the second term is the scalar exchange term, which is obtained from 
the explicit symmetry breaking term, and, the
last three terms ($\sim (\partial_\mu {\bar D})(\partial ^\mu D)$)
are known as the range terms. 
The parameters $d_1$ and $d_2$  in the range terms are fitted 
to the empirical values of kaon-nucleon scattering lengths
for $I=0$ and $I=1$ \cite{isoamss,isoamss2}.
From the interaction Lagrangian density, using the Fourier 
transformations of equations of motion, the dispersion relations 
for $D$ and $\bar{D}$ mesons are obtained as
\begin{equation}
-\omega^{2}+\vec{k}^{2}+m_{D(\bar D)}^{2}
-\Pi_{D(\bar D)}(\omega,\vert\vec{k}\vert) = 0,
\label{dispersion}
\end{equation}
where, $m_{D(\bar D)}$ is the vacuum mass of the $D(\bar D)$ meson and
$\Pi_{D(\bar D)}(\omega,\vert\vec{k}\vert)$ denotes the self-energy 
of the $D\left( \bar{D} \right) $ meson.
For the $D$ 
meson doublet $ \left( D^{0} , D^{+}\right) $, the expression of self energy 
 is
 given as \cite{amarvepja}
\begin{align}
& \Pi _D (\omega, |\vec k|)=  \frac {1}{4 f_D^2}\Big [3 (\rho_p +\rho_n)
\pm (\rho_p -\rho_n) 
+2\Big(\left( \rho_{\Sigma^{+}}+ \rho_{\Sigma^{-}}\right) \pm
\left(\rho_{\Sigma^{+}}- \rho_{\Sigma^{-}}\right) \Big)
+2(\rho_{\Lambda^{0}}+\rho_{\Sigma^{0}}) 
\nonumber\\ &
+( \left( \rho_{\Xi^{0}}+ \rho_{\Xi^{-}}\right) 
\pm \left(\rho_{\Xi^{0}}- \rho_{\Xi^{-}}\right)) 
\Big ] \omega 
+\frac {m_D^2}{2 f_D} (\sigma ' +\sqrt 2 {\zeta_c} ' \pm \delta ')
\nonumber \\ &
+ \Big [- \frac {1}{f_D}
(\sigma ' +\sqrt 2 {\zeta_c} ' \pm \delta ')
+\frac {d_1}{2 f_D ^2} (\rho_s ^p +\rho_s ^n
+\rho_{s}^{\Lambda^{0}}+\rho_{s}^{\Sigma^{+}}+\rho_{s}^{\Sigma^{0}}+\rho_{s}^{\Sigma^{-}}
+\rho_{s}^{\Xi^{0}}+\rho_{s}^{\Xi^{-}})\nonumber \\
&+\frac {d_2}{4 f_D ^2} \Big (({\rho}^p_s +{\rho}^n_s)
\pm   ({\rho}^p_s -{\rho}^n_s)+\frac{1}{3}{\rho}^{\Lambda^{0}}_s
+({\rho}_s^{\Sigma^{+}}+{\rho}_s^{\Sigma^{-}})
\pm    ({\rho}_s^{\Sigma^{+}}-{\rho}_s^{\Sigma^{-}})
+{\rho}_s^{\Sigma^{0}} \Big ) \Big ]
(\omega ^2 - {\vec k}^2).
\label{selfd}
\end{align}
The $\pm$ signs in the above equation refer to the $D^{0}$ and $D^{+}$ mesons, 
respectively. Also, $\sigma^{\prime}\left( = \sigma - \sigma_{0}\right) $, 
$\zeta_{c}^{\prime}\left( = \zeta_{c} - \zeta_{c0}\right)$, and 
$\delta^{\prime}\left(  = \delta -\delta_{0}\right) $ are the 
fluctuations of the scalar-isoscalar fields $\sigma$, $\zeta_{c}$ and
the scalar-isovector field $\delta$ from their 
vacuum expectation values in the strange hyperonic medium. The vacuum expectation value of $\delta$ 
is zero $\left(\delta_{0}=0 \right)$, since a nonzero value for it 
will break the isospin-symmetry of the vacuum. 
Since $\zeta_{c}$ is made of heavy charm quark-antiquark pair, its medium modification is not considered in the present work, i.e, $\zeta_{c}^{\prime} = 0$
is assumed \cite{amarindam,amarvepja}.
For the  $\bar{D}$ meson doublet 
$\left(\bar{D}^{0},D^{-}\right)$, the self-energy is given as 
\begin{align}
& \Pi_{\bar D} (\omega, |\vec k|) = 
 -\frac {1}{4 f_D^2}\Big [3 (\rho_p +\rho_n)
\pm (\rho_p -\rho_n)
 +2\Big(\left( \rho_{\Sigma^{+}}+ \rho_{\Sigma^{-}}\right) \pm
\left(\rho_{\Sigma^{+}}- \rho_{\Sigma^{-}}\right) \Big)
+2(\rho_{\Lambda^{0}}+\rho_{\Sigma^{0}})\nonumber\\
& +( \left( \rho_{\Xi^{0}}+ \rho_{\Xi^{-}}\right) \pm
\left(\rho_{\Xi^{0}}- \rho_{\Xi^{-}}\right))\Big ] \omega
+\frac {m_D^2}{2 f_D} (\sigma ' +\sqrt 2 {\zeta_c} ' \pm \delta ')
\nonumber \\
& + \Big [- \frac {1}{f_D}
(\sigma ' +\sqrt 2 {\zeta_c} ' \pm \delta ') 
+\frac {d_1}{2 f_D ^2} (\rho_s ^p +\rho_s ^n
+\rho_{s}^{\Lambda^{0}}+\rho_{s}^{\Sigma^{+}}
+\rho_{s}^{\Sigma^{0}}+\rho_{s}^{\Sigma^{-}}
+\rho_{s}^{\Xi^{0}}+\rho_{s}^{\Xi^{-}})\nonumber \\
&+\frac {d_2}{4 f_D ^2} \Big (({\rho}^p_s +{\rho}^n_s)
\pm   ({\rho}^p_s -{\rho}^n_s)+\frac{1}{3}{\rho}^{\Lambda^{0}}_s
+({\rho}_s^{\Sigma^{+}}+{\rho}_s^{\Sigma^{-}})
\pm    ({\rho}_s^{\Sigma^{+}}-{\rho}_s^{\Sigma^{-}})
+{\rho}_s^{\Sigma^{0}} \Big ) \Big ]
(\omega ^2 - {\vec k}^2).
\label{selfdbar}
\end{align}
where the $\pm$ sign refers to the $\bar{D}^{0}(D^{-})$ meson. 

\begin{figure}[ht!] 
\vskip -2.7in
 \includegraphics[width=1.\textwidth]{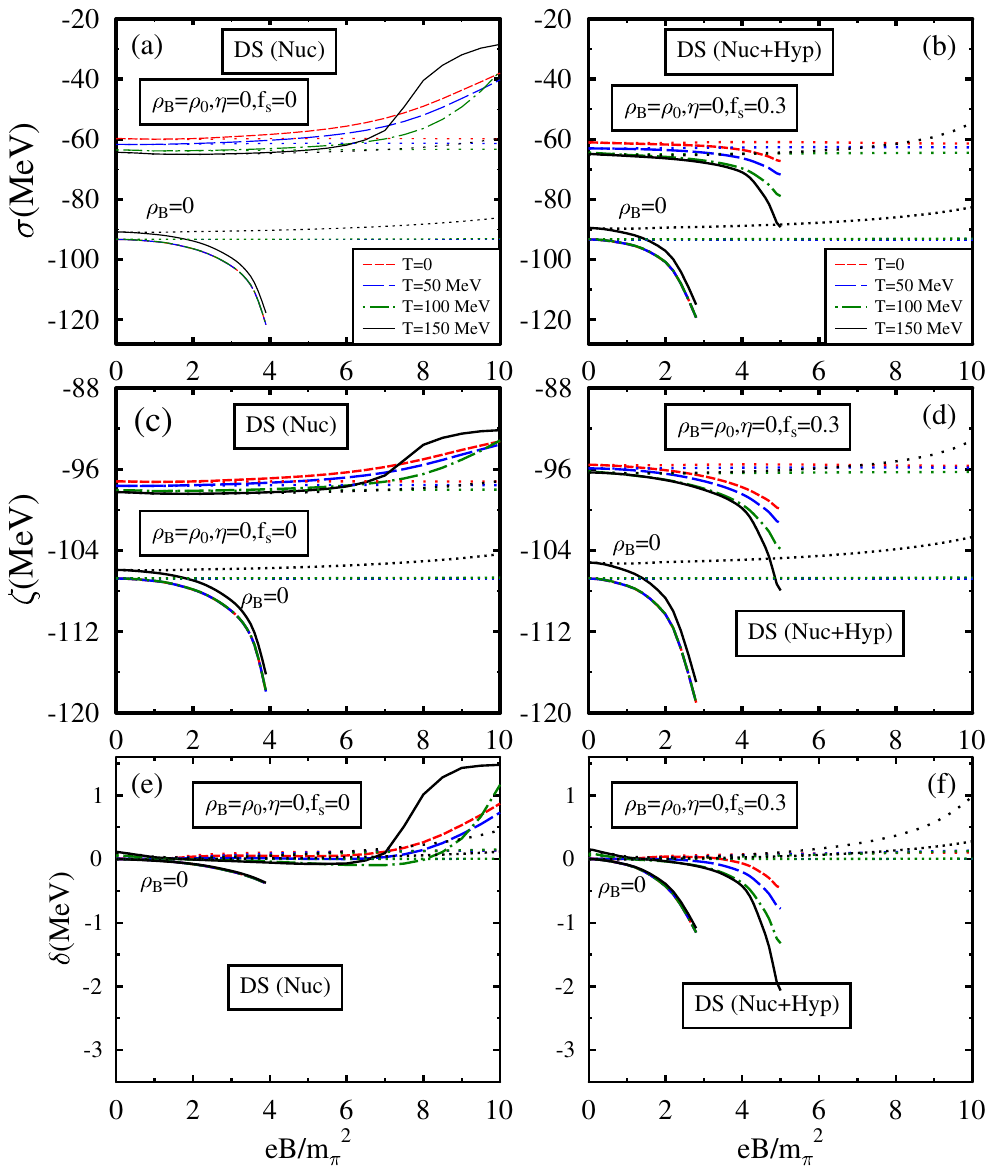}\hfill
\vskip -0.4in
	\caption{The scalar fields $\sigma, \zeta$ and $\delta$ are plotted 
as functions of magnetic field $eB/m_\pi^2$ for $\rho_B=0$ and
$\rho_B=\rho_0$ (for symmetric ($\eta = 0$) nuclear matter ($f_s = 0$) 
(in subplots (a), (c) and (e)) and for hyperonic matter 
(with $f_s$ = 0.3) (in subplots (b), (d) and (f))) 
for different temperatures, 
considering baryonic Dirac sea (DS) contributions. These 
are also plotted when the DS contributions are ignored
(the closely and widely spaced dotted lines
for $\rho_B=0$ and $\rho_B=\rho_0$).}
	\label{sg_zeta_dlt_eta0_rhb0_zero_density}
\end{figure}
\begin{figure}[ht!] 
\vskip -2.7in
 \includegraphics[width=1.\textwidth]{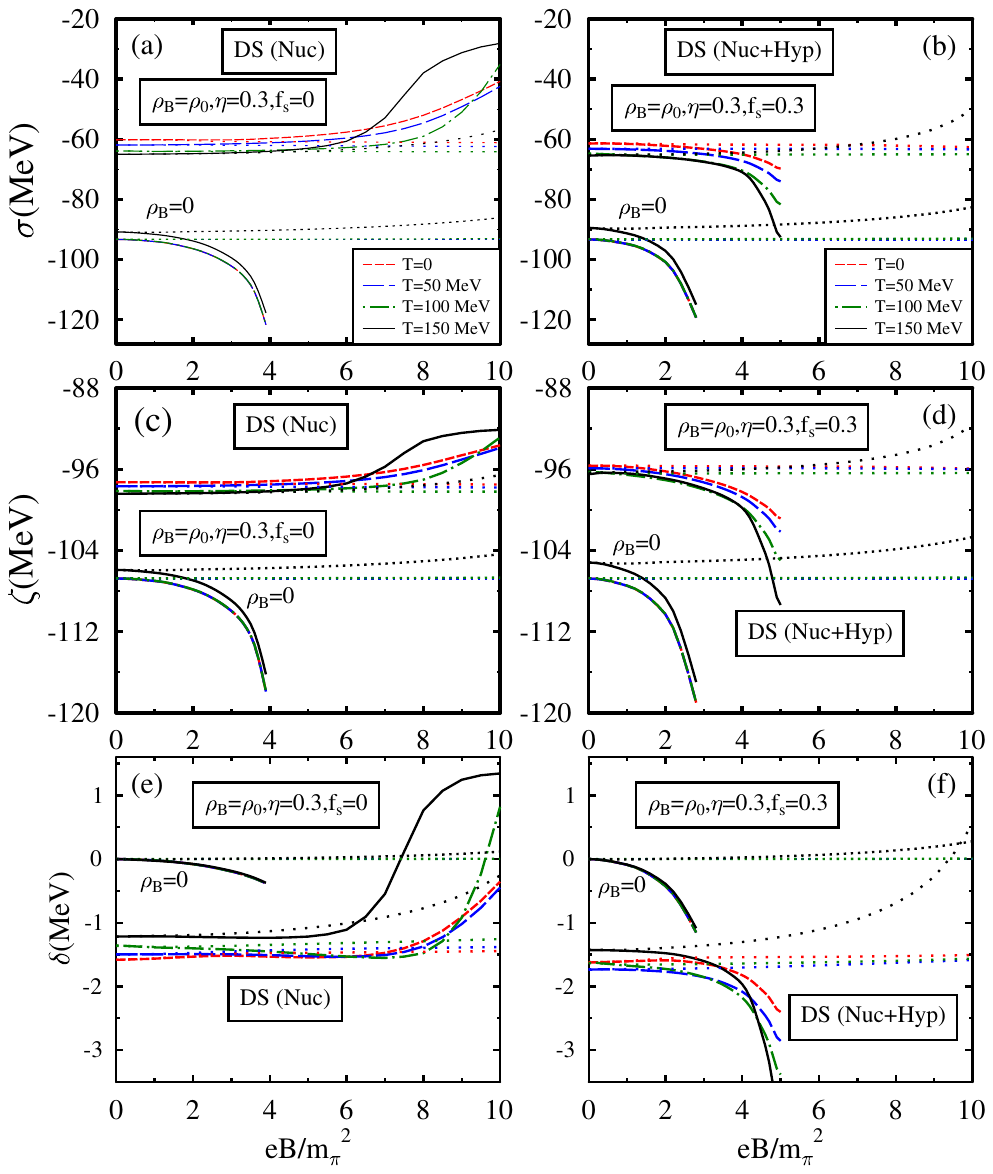}\hfill
\vskip -0.4in
	\caption{Same as figure \ref{sg_zeta_dlt_eta0_rhb0_zero_density},
for asymmetric matter ($\eta$=0.3).}
	\label{sg_zeta_dlt_eta3_rhb0_zero_density}
\end{figure}
\begin{figure}[ht!] 
\vskip -2.7in
 \includegraphics[width=1.\textwidth]{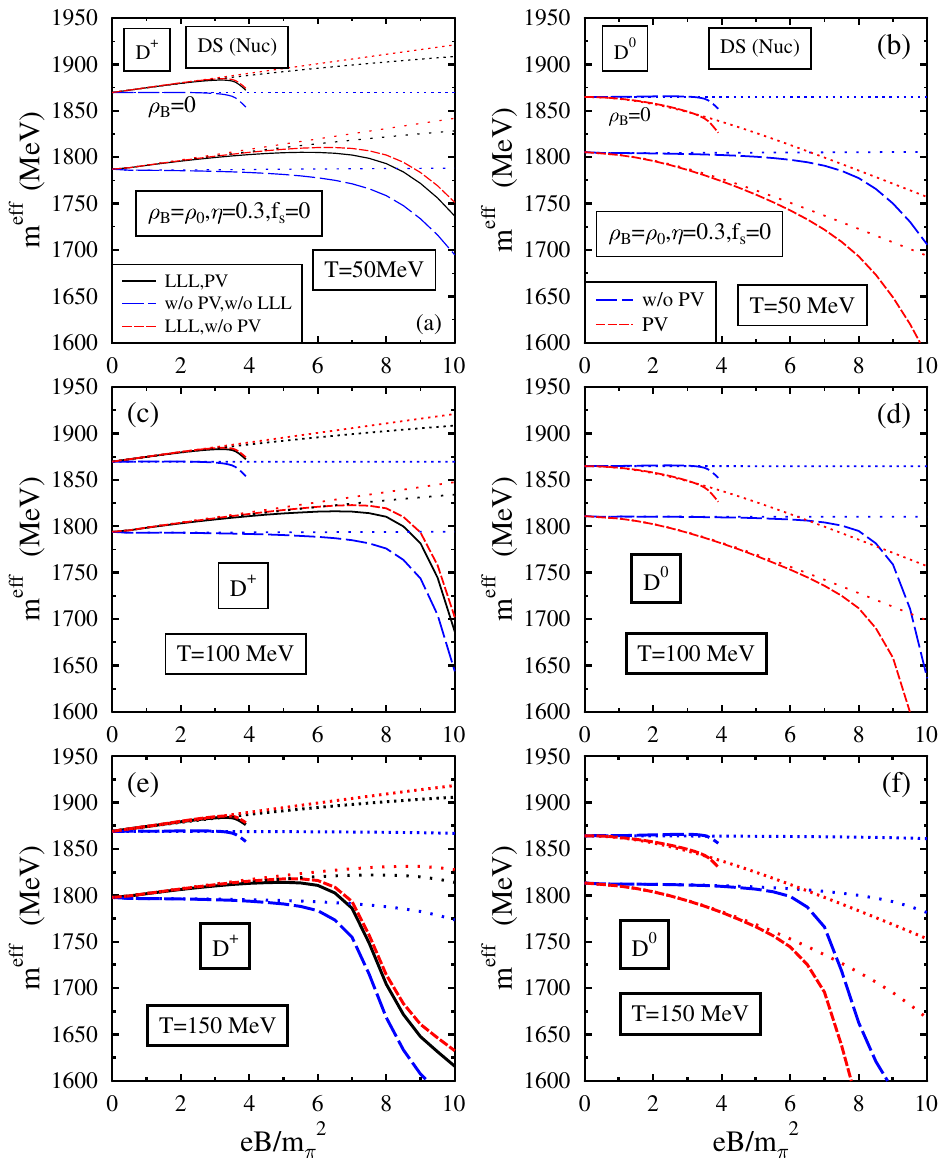}\hfill
\vskip -0.4in
	\caption{The masses of pseudoscalar $D^{+}$ 
(in subplots (a), (c) and (e)) 
and $D^{0}$ 
(in subplots (b), (d) and (f))
 mesons 
are plotted as functions of magnetic field $eB/m_\pi^2$,
at T=50, 100 and 150 MeV respectively, 
for $\rho_B=0$
and for asymmetric nuclear matter ($\rho_B=\rho_0$, $\eta = 0.3$ and $f_s =0$), 
considering baryonic Dirac sea contributions,
with and without PV mixing. These are also plotted
when Dirac sea contributions are neglected 
(the closely and widely spaced dotted lines
for $\rho_B=0$ and $\rho_B=\rho_0$).}
	\label{md_eta3_fs0_Temp_rhb0_zero_density}
\end{figure}
\begin{figure}[ht!] 
\vskip -2.7in
 \includegraphics[width=1.\textwidth]{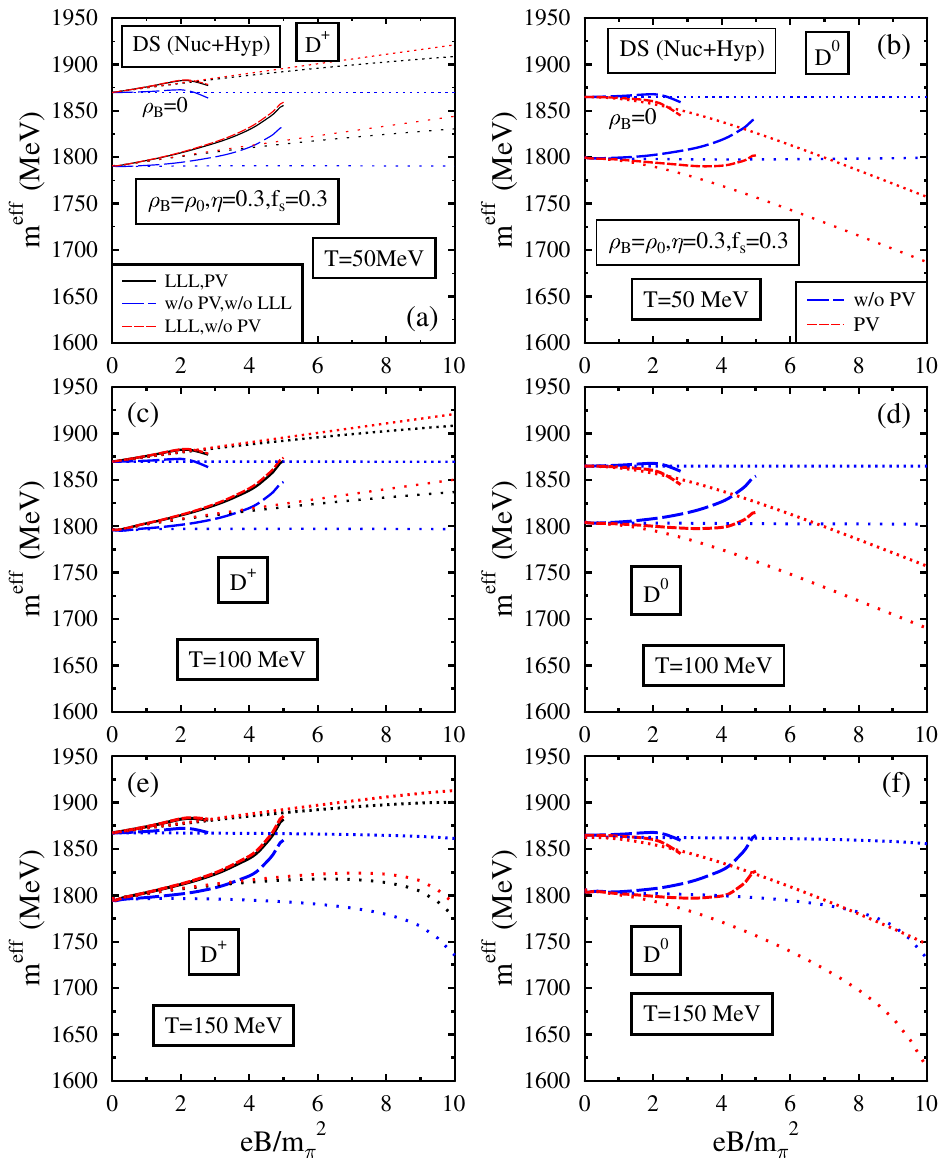}\hfill
\vskip -0.4in
	\caption{Same as figure \ref{md_eta3_fs0_Temp_rhb0_zero_density},
for hyperonic matter (with $f_s$=0.3).}
	\label{md_eta3_fs3_Temp_rhb0_zero_density}
\end{figure}
\begin{figure}[ht!] 
\vskip -2.7in
 \includegraphics[width=1.\textwidth]{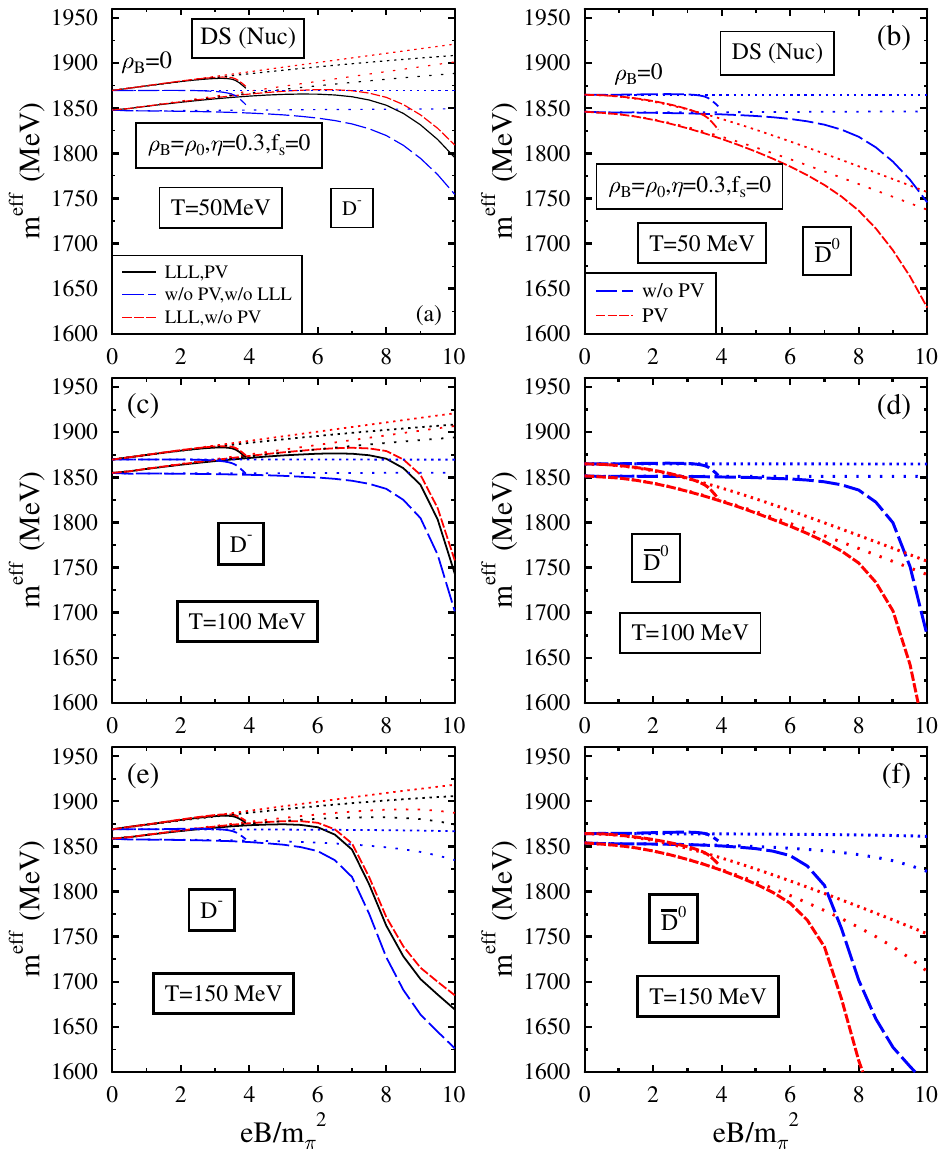}\hfill
\vskip -0.4in
	\caption{The masses of pseudoscalar $D^{-}$ 
(in subplots (a), (c) and (e)) 
and $\bar{D^{0}}$
(in subplots (b), (d) and (f))
 mesons 
are plotted as functions of magnetic field $eB/m_\pi^2$,
at T=50, 100 and 150 MeV respectively, 
for $\rho_B=0$
and for asymmetric nuclear matter ($\rho_B=\rho_0$, $\eta = 0.3$ and $f_s =0$), 
considering baryonic Dirac sea contributions,
with and without PV mixing. These are also plotted
when Dirac sea contributions are neglected 
(the closely and widely spaced dotted lines
for $\rho_B=0$ and $\rho_B=\rho_0$).}
	\label{mdbar_eta3_fs0_Temp_rhb0_zero_density}
\end{figure}
\begin{figure}[ht!] 
\vskip -2.7in
 \includegraphics[width=1.\textwidth]{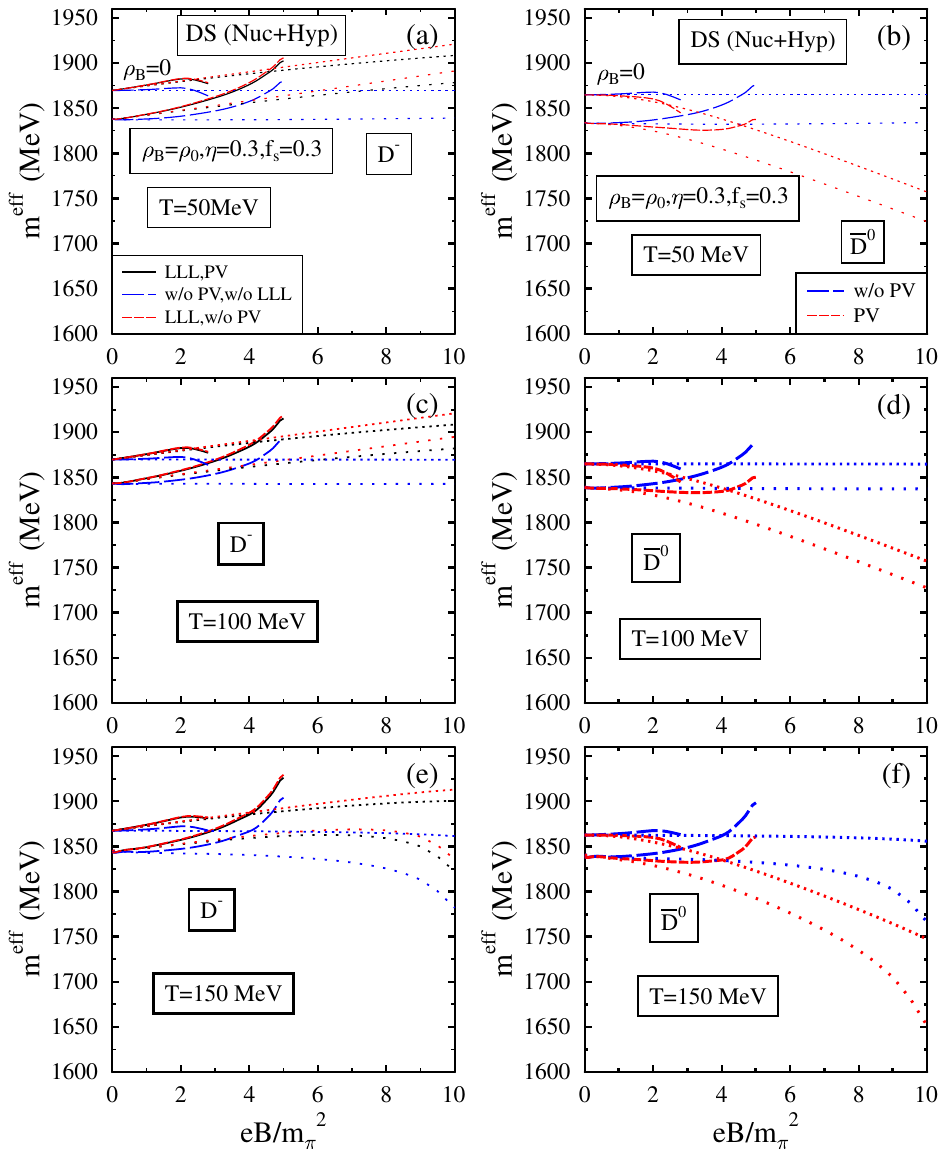}\hfill
\vskip -0.4in
	\caption{Same as figure \ref{mdbar_eta3_fs0_Temp_rhb0_zero_density},
for hyperonic matter (with $f_s$=0.3).}
	\label{mdbar_eta3_fs3_Temp_rhb0_zero_density}
\end{figure}
\begin{figure}[ht!] 
\vskip -2.7in
 \includegraphics[width=1.\textwidth]{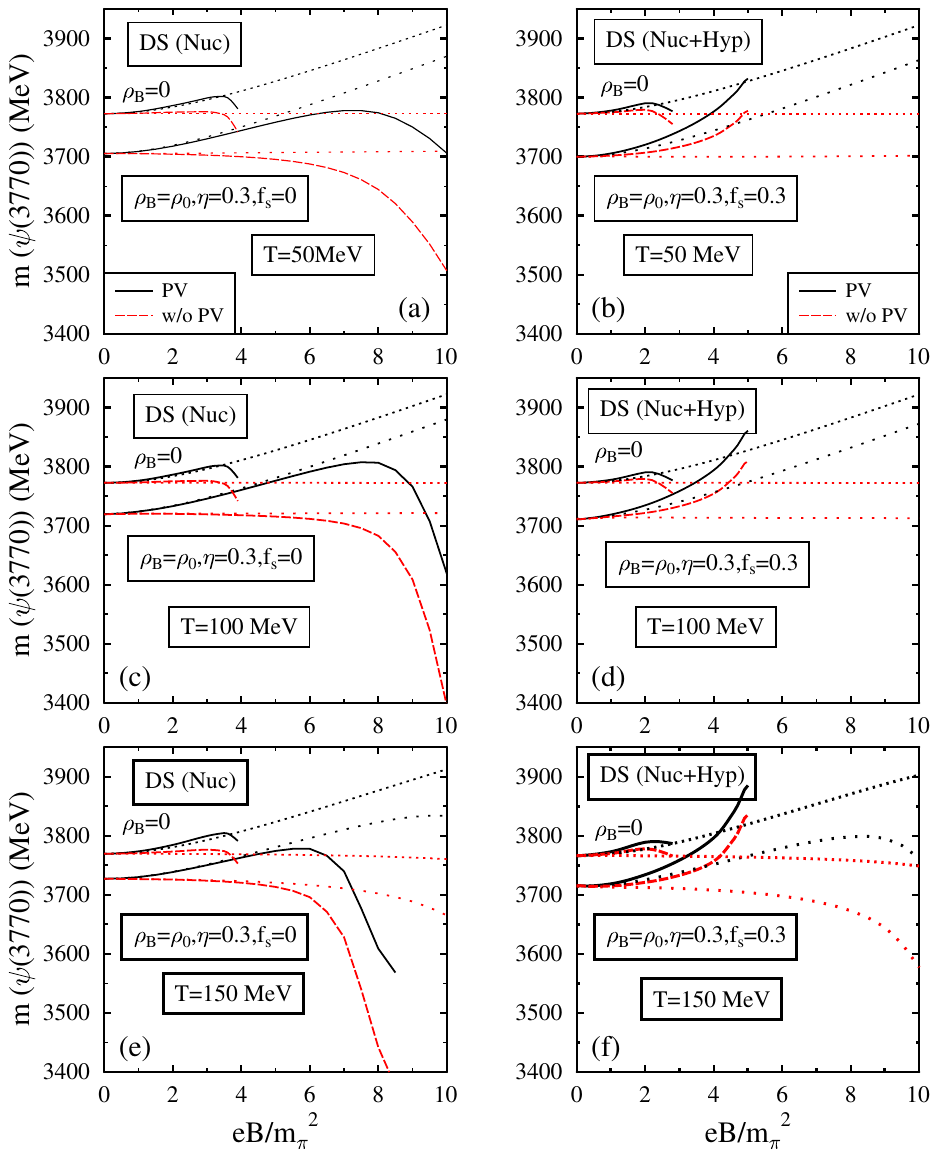}\hfill
\vskip -0.4in
	\caption{The mass of charmonium $\psi(3770)$ is plotted as 
a function of magnetic field $eB/m_\pi^2$ for $\rho_B=0$ and
for asymmetric ($\eta = 0.3$)
nuclear matter ($f_s$=0) 
(in subplots (a), (c) and (e)) 
and strange hadronic matter
(with $f_s = 0.3$)
(in subplots (b), (d) and (f))
at T=50, 100 and 150 MeV respectively, 
for $\rho_B = \rho_0$, with and without PV mixing,
considering baryonic Dirac sea contributions. 
The masses are also plotted when Dirac sea contributions are neglected 
(the closely and widely spaced dotted lines
for $\rho_B=0$ and $\rho_B=\rho_0$).}
	\label{psi3770_eta3_Temp_rhb0_zero_density}
\end{figure}
\begin{figure}[ht!] 
\vskip -2.7in
 \includegraphics[width=1.\textwidth]{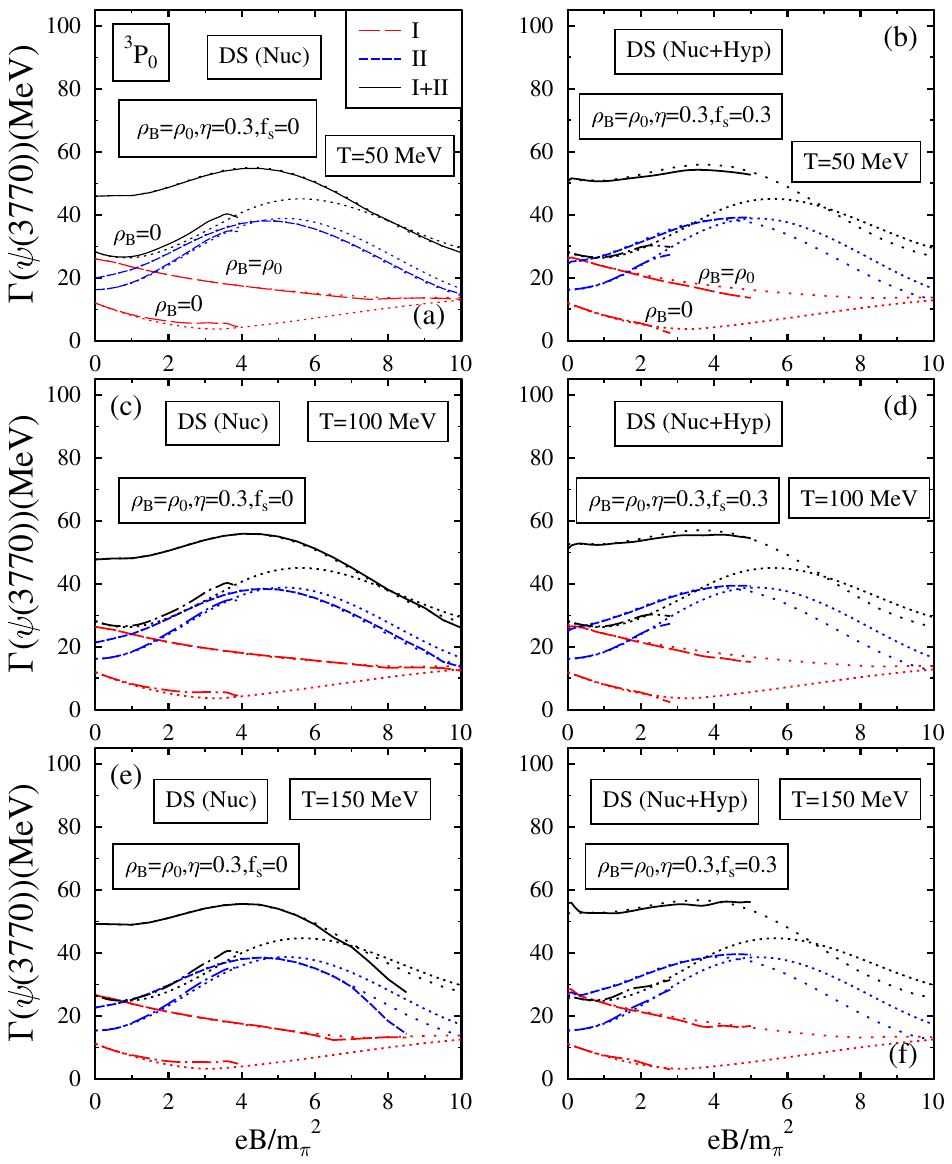}\hfill
\vskip -0.4in
	\caption{Decay widths of charmonium $\psi(3770)$ 
to (I) $D^+D^-$, (II) $D^0\bar{D^0}$
		and (III) total $D\bar{D}$, i.e., (I+II) are plotted 
as functions of magnetic field $eB/m_\pi^2$ for $\rho_B=0$ and
for asymmetric ($\eta =0.3$) nuclear matter ($f_s = 0$)
(in subplots (a), (c) and (e)) 
 and strange 
hadronic matter ($f_s$=0.3)
(in subplots (b), (d) and (f))
at T=50, 100 and 150 MeV respectively, 
for $\rho_B = \rho_0$, 
using the $^3P_0$ model, considering baryonic Dirac sea 
contributions and PV mixing. 
These are also plotted when Dirac sea contributions are neglected 
(the closely and widely spaced dotted lines
for $\rho_B=0$ and $\rho_B=\rho_0$).}
	\label{Decaywidth_psi3770_3p0_eta3_Temp_rhb0_zero_density}
\end{figure}
\begin{figure}[ht!] 
\vskip -2.7in
 \includegraphics[width=1.\textwidth]{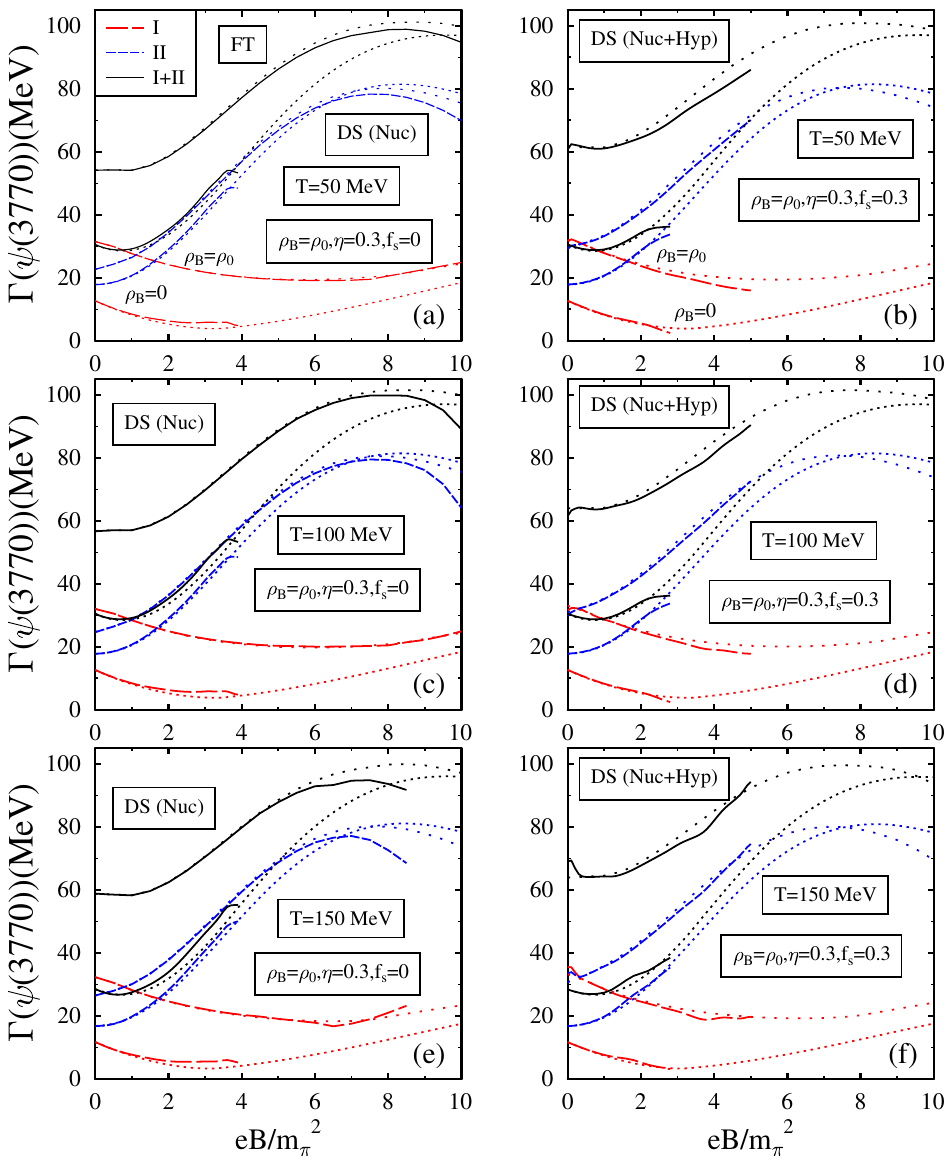}\hfill
\vskip -0.4in
	\caption{Same as Fig. 
\ref{Decaywidth_psi3770_3p0_eta3_Temp_rhb0_zero_density},
using the field theoretical (FT) model of composite hadrons.}
	\label{Decaywidth_psi3770_FT_eta3_Temp_rhb0_zero_density}
\end{figure}
\begin{figure}
\vskip -2.7in
    \includegraphics[width=0.9\textwidth]{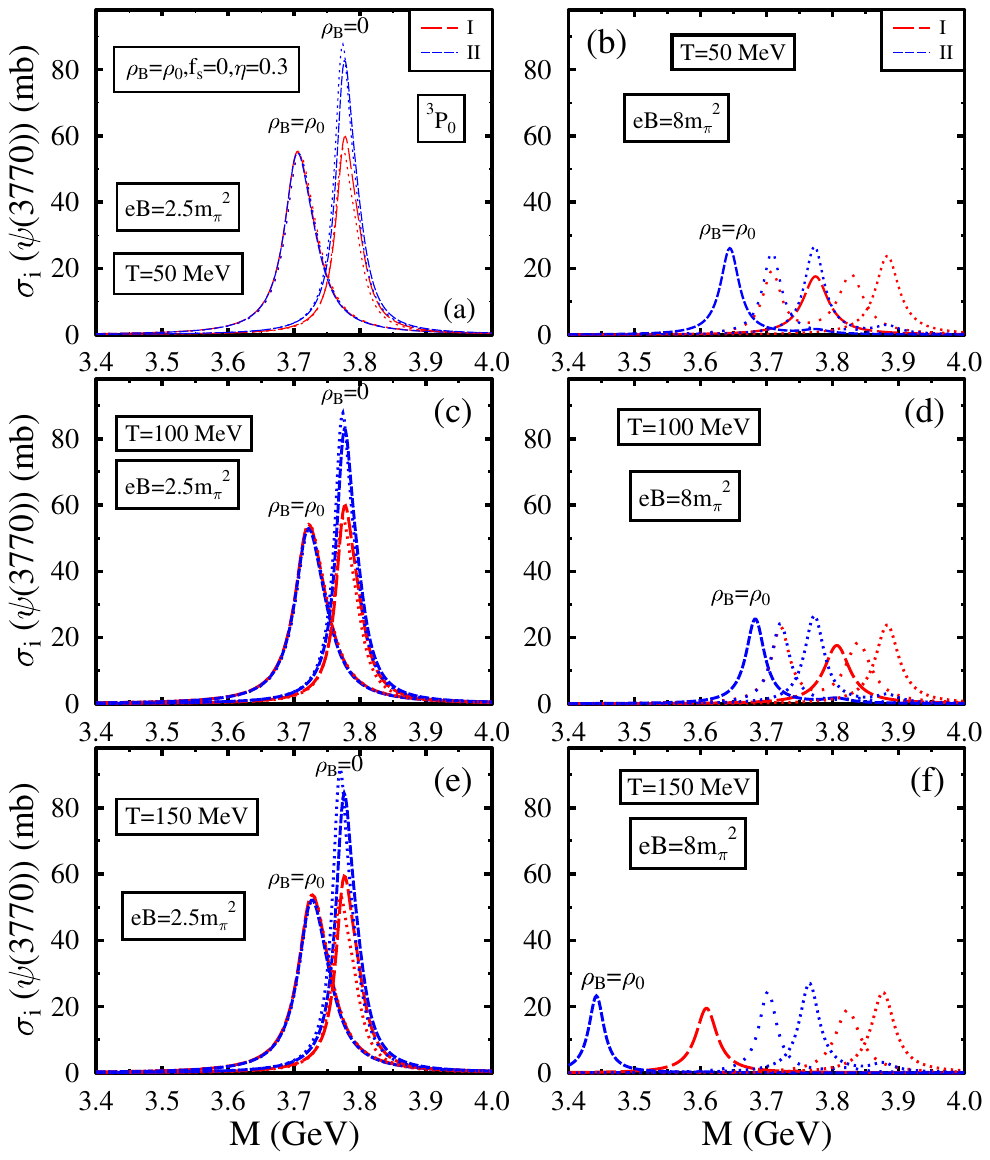}
\vskip -0.4in
    \caption{Production cross-sections of $\psi(3770)$
due to scattering of (I) $D^+D^-$ and (II) $D^0\bar{D^0}$ mesons 
are plotted
as functions of the invariant mass,
for $eB=2.5 m_\pi^2$
(in subplots (a), (c) and (e)) 
and $eB=8 m_\pi^2$
(in subplots (b), (d) and (f))
at T=50, 100 and 150 MeV respectively, 
 for $\rho_B=0$ 
as well as for asymmetric ($\eta$=0.3) nuclear ($f_s$=0) 
matter at $\rho_B=\rho_0$, with the decay widths obtained 
using the $^3P_0$ model.
These are also plotted when Dirac sea contributions are neglected 
(the closely and widely spaced dotted lines
for $\rho_B=0$ and $\rho_B=\rho_0$).}
    \label{REV2_Cross_section_3p0_psi3770_eta3_fs0_2_5_8mpi2}
\end{figure}
\begin{figure}
\vskip -2.7in
    \includegraphics[width=0.9\textwidth]{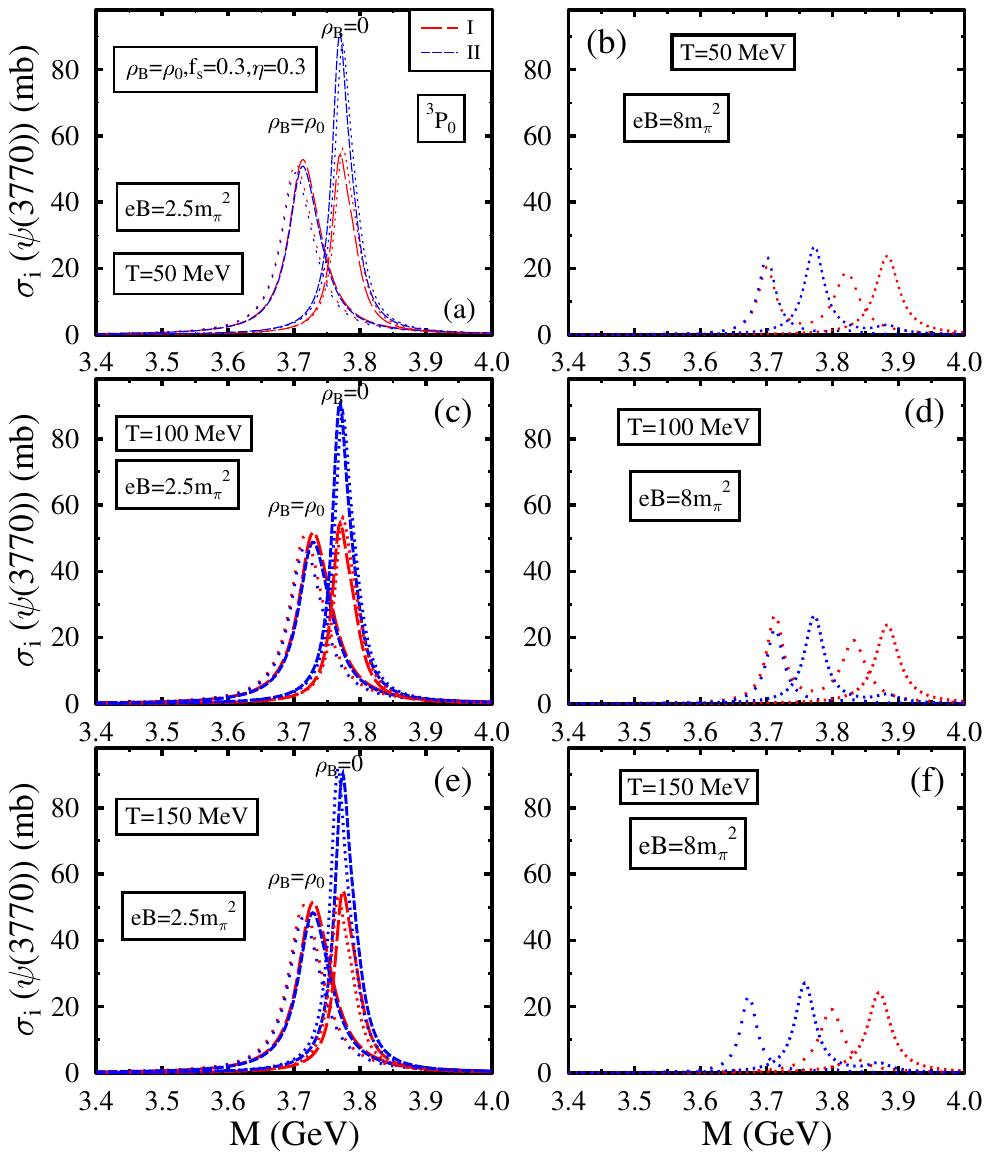}
\vskip -0.4in
    \caption{Same as Fig. 
\ref{REV2_Cross_section_3p0_psi3770_eta3_fs0_2_5_8mpi2},
for strange hadronic matter (with $f_s$=0.3).}
    \label{REV2_Cross_section_3p0_psi3770_eta3_fs3_2_5_8mpi2}
\end{figure}
\begin{figure}
\vskip -2.7in
    \includegraphics[width=0.9\textwidth]{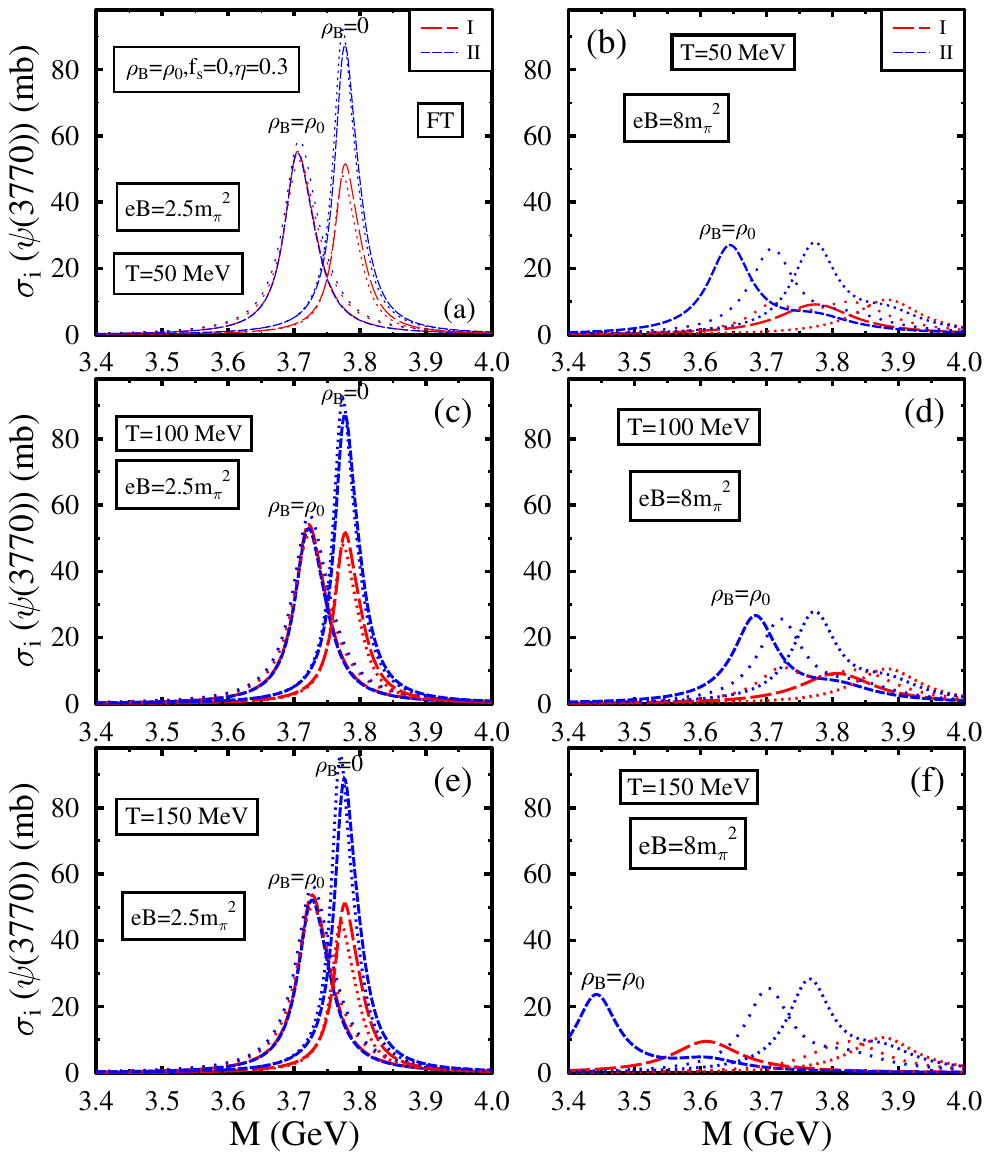}
\vskip -0.4in
    \caption{Same as Fig. 
\ref{REV2_Cross_section_3p0_psi3770_eta3_fs0_2_5_8mpi2},
with decay widths calculated using the field theoretical (FT)
model of composite hadrons.}
    \label{REV2_Cross_section_FT_psi3770_eta3_fs0_2_5_8mpi2}
\end{figure}
\begin{figure}
\vskip -2.7in
    \includegraphics[width=0.9\textwidth]{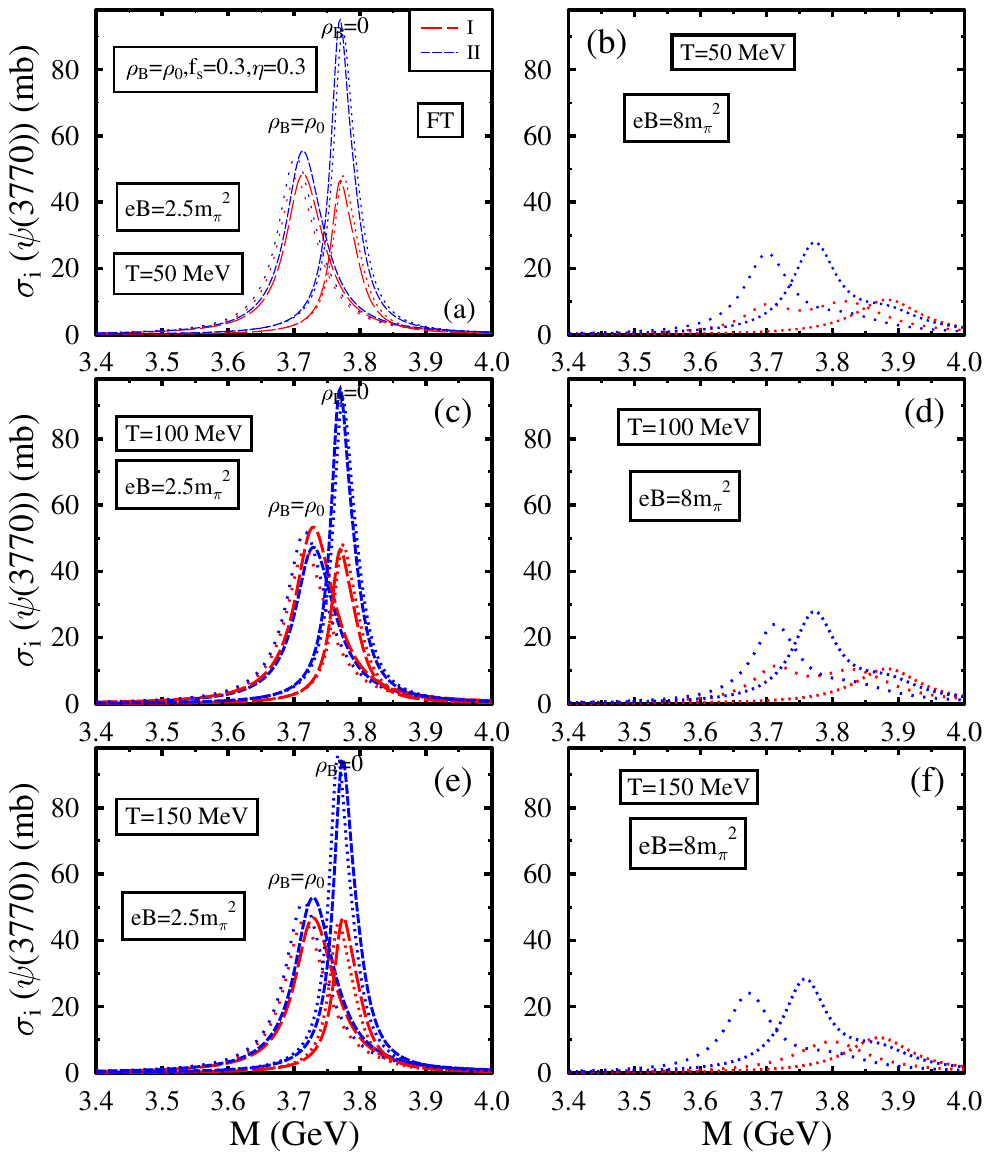}
\vskip -0.4in
    \caption{Same as Fig. 
\ref{REV2_Cross_section_3p0_psi3770_eta3_fs3_2_5_8mpi2},
with decay widths calculated using the field theoretical (FT)
model of composite hadrons.}
    \label{REV2_Cross_section_FT_psi3770_eta3_fs3_2_5_8mpi2}
\end{figure}
The in-medium mass $m_{D(\bar{D})}^{*}$ for $D(\bar{D})$ meson 
is obtained by solving the dispersion relation given by
equation (\ref{dispersion}) for momentum $|\vec{k}|=0$. 
The medium modifications of the masses
of these open charm mesons 
arise due to their interactions with the scalar mesons
and baryons (through their number and scalar densities). 
It may be noted here that there is no contribution due to
interaction with the pions, since the expectation values
of these odd-parity fields are zero in the mean field approximation.
For the charged $D$  ($D^{\pm}$) mesons,
additional contributions 
from the lowest Landau levels (LLL) are taken into account,
and the effective mass of these mesons are given as
$m^{eff}_{D^{\pm}}=\sqrt{m_{D^{\pm}}^{*2} + |eB|}$
\cite{open_charm_mag_AM_SPM}.
 
The effective mass of charmonium state $\psi(3770)$ is obtained from the medium 
modification of the dilaton field, $\chi$, which simulates the gluon condensates
of QCD within the chiral effective model. 
The mass shift of the charmonium states 
arises from the medium modification
of the scalar gluon condensate in the leading order 
and is given as \cite{pes1,pes2,voloshin,leeko} 
\begin{eqnarray}
\Delta m_{\Psi} &=& \frac{1}{18} \int d{|{\bf k}|}^{2} \left\langle 
\left\vert \frac{\partial \psi ({\bf k})}{\partial {\bf k}} \right\vert^{2} \right\rangle 
\frac{|{\bf k}|}{|{\bf k}|^{2} 
/ m_{c} + \epsilon} 
\bigg ( 
\left\langle \frac{\alpha_{s}}{\pi} 
G_{\mu\nu}^{a} G^{\mu\nu a}\right\rangle -
\left\langle \frac{\alpha_{s}}{\pi} 
G_{\mu\nu}^{a} G^{\mu\nu a}\right\rangle _{0}
\bigg ),
\label{mass1}
\end{eqnarray}
where
\begin{equation}
\left\langle \left\vert \frac{\partial \psi (\bf k)}{\partial {\bf k}}
\right\vert^{2} \right\rangle
=\frac {1}{4\pi}\int
\left\vert \frac{\partial \psi (\bf k)}{\partial {\bf k}} \right\vert^{2}
d\Omega.
\end{equation}
Equating the trace of the energy momentum tensor of the chiral model
and QCD relates the scalar gluon condensates to the dilaton field,
$\chi$ as \cite{amarvepja}
\begin{equation}
\left\langle \frac{\alpha_{s}}{\pi} 
G_{\mu\nu}^{a} G^{\mu\nu a} \right\rangle 
= \frac{24}{(33-2N_f)}(1-d)\chi^{4}, 
\label{chiglu}
\end{equation} 
in the limiting situation of massless quarks
in the energy momentum tensor of QCD.

Using equations (\ref{mass1}) and (\ref{chiglu}), one obtains
the mass shift of the charmonium state as
\cite{amarvdmesonTprc,amarvepja}
\begin{equation}
\Delta m_{\Psi}= \frac{4}{81} (1 - d) \int d |{\bf k}|^{2}
\left\langle \left\vert \frac{\partial \psi (\bf k)}{\partial {\bf k}}
\right\vert^{2} \right\rangle
\frac{|{\bf k}|}{|{\bf k}|^{2} / m_{c} + \epsilon}
 \left( \chi^{4} - {\chi_0}^{4}\right).
\label{mass_shift_psi3770}
\end{equation}
In equation (\ref{mass_shift_psi3770}), $d$ is a parameter introduced
in the scale breaking term in the Lagrangian, $\chi$ and $\chi_0$
are the values of the dilaton field in the magnetized medium 
and in vacuum respectively. 
The wave functions of the charmonium states,
$\psi(\bf k)$ are assumed to be harmonic oscillator
wave functions, $m_{c}$ is the mass of the charm
quark, $\epsilon=2m_{c}-m_{\psi}$
is the binding energy of the charmonium 
state of mass, $m_{\psi}$.\\

\newpage
\noindent {\bf {Pseudoscalar-Vector meson (PV) mixing:}}\\
In the presence of a magnetic field, there is mixing between
the spin 0 (pseudoscalar) meson and spin 1 (vector) mesons, which modifies
the masses of these mesons 
\cite{charmonium_mag_QSR,charmonium_mag_lee,Suzuki_Lee_2017,Alford_Strickland_2013,charmdw_mag,open_charm_mag_AM_SPM,strange_AM_SPM,Quarkonia_B_Iwasaki_Oka_Suzuki}.
The mass modifications have been studied 
using a phenomenological Lagrangian density of the form 
\cite{charmonium_mag_lee,Suzuki_Lee_2017,Quarkonia_B_Iwasaki_Oka_Suzuki} 
\begin{equation}
{\cal L}_{PV\gamma}=\frac{g_{PV}}{m_{av}} e {\tilde F}_{\mu \nu}
(\partial ^\mu P) V^\nu,
\label{PVgamma}
\end{equation}
for the heavy quarkonia \cite{charmonium_mag_QSR,charmonium_mag_lee,Suzuki_Lee_2017,charmdw_mag,upslndw_mag},
the open charm mesons \cite{open_charm_mag_AM_SPM}
and the strange ($K$ and $\bar K$) mesons
\cite{strange_AM_SPM}.
In equation (\ref{PVgamma}), $m_{av}=(m_V+m_P)/2$, 
$m_P$ and $m_V$ are the   masses 
for the pseudoscalar and vector charmonium states,
${\tilde F}_{\mu \nu}$ is the dual electromagnetic field.
The coupling parameter $g_{PV}$
is fitted from the observed value of the radiative decay width, 
$\Gamma(V\rightarrow P+\gamma)$.
Assuming the spatial momenta of the heavy quarkonia to be zero,
there is observed to be mixing between the pseudoscalar and the 
longitudinal component of the vector field from their 
equations of motion obtained with the phenomenological
$PV\gamma$ interaction given by equation (\ref{PVgamma}).
The physical masses of the pseudoscalar and the longitudinal component
of the vector mesons including the mixing effects,
obtained by solving their equations of motion, are given as
\cite{charmonium_mag_lee,Suzuki_Lee_2017,Quarkonia_B_Iwasaki_Oka_Suzuki} 
\begin{equation}
m^{(PV)}_{P(V^{||})}=\frac{1}{2} \Bigg ( M_+^2
+\frac{c_{PV}^2}{m_{av}^2} \mp
\sqrt {M_-^4+\frac{2c_{PV}^2 M_+^2}{m_{av}^2}
+\frac{c_{PV}^4}{m_{av}^4}} \Bigg),
\label{mpv_long}
\end{equation}
where $M_+^2=m_P^2+m_V^2$, $M_-^2=m_V^2-m_P^2$ and
$c_{PV}= g_{PV} eB$.
In the above equation, the `$\mp$' corresponds to the
mass of the pseudoscalar (longitudinal component
of the vector) meson.
The effective Lagrangian term given by equation
(\ref{PVgamma}) has been observed to lead to the
mass modifications of the longitudinal $J/\psi$ and
$\eta_c$ due to the presence of the magnetic field,
which agree extermely well with
a study of these charmonium states using a
QCD sum rule approach incorporating the mixing effects
\cite{charmonium_mag_QSR,charmonium_mag_lee}.

The PV mixing effects for the open charm mesons 
(due to $D-D^*$ and $\bar D-\bar D^*$ mixings)
\cite{open_charm_mag_AM_SPM}, in addition to the mixing 
of the charmonium states (due to $J/\psi-\eta_c$, $\psi'-\eta_c'$ and
$\psi(3770)-\eta_c'$ mixings)
\cite{charmdw_mag,open_charm_mag_AM_SPM}, 
as calculated using the phenomenological Lagrangian given by 
equation (\ref{PVgamma}) have been observed to lead
to appreciable drop (rise) in the mass of the pseudoscalar
(longitudinal component of the vector) meson. These were
observed to modify the partial decay width of 
$\psi(3770)\rightarrow D\bar D$ 
\cite{charmdw_mag,open_charm_mag_AM_SPM}, with the
modifications being much more dominant due to the 
PV mixing in the open charm ($D-{D^*}$ and ${\bar D}-{\bar D}^*$)
mesons. 

\subsection{Partial Decay width of Charmonium state to $D\bar D$}

In this subsection, we briefly describe the study of
partial decay width of charmonium state, $\psi(3770)$ 
to open charm ($D$ and $\bar D$) mesons, in hot asymmetric 
strange hadronic matter in the prsence of a magnetic field. 
These are studied using (I) the $^3P_0$ model, and,
(II) a field theoretical (FT) model of composite hadrons with quark
(and antiquark) constituents. 
In both of these models, the decay proceeds with creation
of a light quark-antiquark pair and the heavy charm quark 
(antiquark) of the parent charmonium state
combining with the light antiquark (quark) to produce the $D\bar D$
in the final state. 

For the charmonium state decaying at rest,
the in-medium decay widths are obtained in terms of the 
magntitude of the 3-momentum of the outgoing $D$ ($\bar D$)
given as 
\begin{equation}
|{\bf p}|=\Bigg (\frac{{M_\psi}^2}{4}-\frac {{m_D}^2+{m_{\bar D}}^2}{2}
+\frac {({m_D}^2-{m_{\bar D}}^2)^2}{4 {M_\psi}^2}\Bigg)^{1/2},
\label{pd}
\end{equation}
where, $M_\psi$, $m_D$ and $m_{\bar D}$ are the in-medium masses of the 
charmonium state, $\psi(3770)$, $D$ and $\bar D$ mesons.
The mass modifications of the charmonium states
and the open charm mesons are observed to lead
to substantial modification of the partial decay width 
of $\psi(3770)\rightarrow D\bar D$ \cite{charmdw_mag}
due to $\psi(3770)-\eta^{'}_c$ mixing, as well as, due to
$D(\bar D)-D^* (\bar D^*)$ mixing effects 
\cite{open_charm_mag_AM_SPM,HQ_DW_DS_AM_SPM_2023}. 
The medium modified masses of $D$, $\bar{D}$ mesons and the charmonium 
state $\psi(3770)$ in magnetized isospin asymmetric strange hadronic matter
at finite temperatures, and, their effect on the decay width of
$\psi(3770)\rightarrow D\bar D$ are calculated in the present 
investigation. The masses of the charmonium and open charm mesons 
are computed including the effects of the Dirac sea as well as PV mixing,
with additional contributions from the lowest Landau level (LLL) 
for the charged $D^\pm$ mesons, as has been described in the previous
subsections.

The $^3P_0$ model \cite{friman,micu,yaouanc,barnesclose}
describes the charmonium decaying to $D\bar D$ with creation of 
a light quark-antiquark pair in the $^3P_0$ state, 
where the light quark and antiquark
combine with  the heavy charm antiquark and quark  of the parent
meson, to produce the $D$ and $\bar D$ mesons in the final state.
The wave functions of the charmonium state and the open charm
mesons are assumed to be of harmonic oscillator type.
The decay width of the charmonium state $\psi(3770)$,
corresponding to the $1D$ state, decaying at rest
to $D\bar D$ is given as 
\cite{barnesclose,friman}
\begin{eqnarray}
&&\Gamma ^{[^3P_0]} (\psi (3770) ({\bf 0}) \rightarrow D ({\bf p})\bar D (-{\bf p}))
= \pi^{1/2} \frac {E_D (|{\bf p}|) E_{\bar D} 
(|{\bf p}|)\gamma^2}{2 M_{\psi}}
\frac { 2^{11} 5}{3^2} \Big (\frac{r}{1+2r^2} \Big )^7
\nonumber \\ &\times & x^3
\Bigg (1-\frac {1+r^2}{5(1+2r^2)}x^2\Bigg)^2
\exp\Big(-\frac {x^2}{2(1+2r^2)}\Big),
\label{dwpsi3770}
\end{eqnarray}
where, $x=|{\bf p}|/\beta_D$, $r=\frac{\beta}{\beta_D}$
is the ratio of the harmonic oscillator strengths of the decaying 
charmonium state and the produced $D(\bar D)$-mesons, 
$E_D (|{\bf p}|)$ and $E_{\bar D} (|{\bf p}|)$ is the energy 
of the outgoing $D$ ($\bar D$) meson given as 
$E_{D(\bar D)} (|{\bf p}|)=(|{\bf p}|^2+{m_{D(\bar D)}}^2)^{1/2}$,
and, $\gamma$ is a measure of 
the strength of the $^3P_0$ vertex \cite{friman,barnesclose}
fitted from the observed decay width of $\psi(3770)\rightarrow D\bar D$
\cite{amarvepja}. 

We next briefly descibe the field theoretical 
model of composite hadrons \cite{spm781,spm782}
used to compute the partial decay widths of 
$\psi(3770)\rightarrow D\bar D$
in magnetized hot isospin asymmetric hyperonic matter.
The model describes the hadrons as comprising of 
quark (and antiquark) constituents. 
The constituent quark field operators of the hadron in motion
are constructed from the constituent quark field operators of
the hadron at rest, by a Lorentz boosting.
Similar to the MIT bag model \cite{MIT_bag}, 
where the quarks (antiquarks) occupy
specific energy levels inside the hadron, it is assumed 
in the present model for the composite hadrons that 
the quark (antiquark) constituents carry fractions
of the mass (energy) of the hadron at rest (in motion) 
\cite{spm781,spm782}.
With explicit constructions of the charmonium
state and the open charm ($D$ and $\bar D$) mesons, 
the charmonium decay width
is calculated using the light quark antiquark pair creation
term of the free Dirac Hamiltonian 
given as
\begin{equation}
{\cal H}_{q^\dagger\tilde q}({\bf x},t=0)
=Q_{q}^{(p)}({\bf x})^\dagger (-i 
{\vec {\alpha}\cdot {\vec \bigtriangledown}} 
+\beta M_q)
{\tilde Q}_q^{(p')}({\bf x}) 
\label{hint}
\end{equation}
where, 
$\vec \alpha = \left (
\begin{array}{cc}  0 & \vec \sigma \\ \vec \sigma & 0
\end{array}\right)$ and
$\beta = \left (
\begin{array}{cc}  I & 0  \\ 0 & -I
\end{array}\right)$ are the Dirac matrices,
$M_q$ is the constituent mass of the light quark (antiquark). 
The subscript $q$ of the field operators in equation (\ref{hint})  
refers to the fact that the light antiquark, $\bar q$ and 
light quark, $q$ are the constituents 
of the $D$ and $\bar D$ mesons with momenta ${\bf p}$ and
${\bf p'}$ respectively in the final state of the
decay of the charmonium state, $\Psi(3770)$.
The decay width of $\psi(3770)\rightarrow D\bar D$
is calculated from the matrix element of 
the light quark-antiquark pair creation part of the free Dirac Hamiltonian,
between the initial and the final state mesons
as given by
\begin{eqnarray}
\langle D ({\bf p}) | \langle {\bar D} ({\bf p'})|
{\int {{\cal H}_{q^\dagger\tilde q}({\bf x},t=0)d{\bf x}}}
|\psi(3770)^m (\vec 0) \rangle 
= \delta({\bf p}+{\bf p}')A_{\psi} (|{\bf p}|)p_m,
\label{tfi}
\end{eqnarray}
which yields the expression for the decay width as given by 
\begin{eqnarray}
\Gamma ^{[FT]}(\psi(3770)\rightarrow D({\bf p}) {\bar D} (-{\bf p}))
= \gamma_\psi^2\frac{8\pi^2}{3}|{\bf p}|^3
\frac {{E_D}(|{\bf p}|) E_{\bar D}(|{\bf p}|)}{M_{\psi}}
A_{\psi}(|{\bf p}|)^2
\label{gammapsiddbar}
\end{eqnarray}
where, $E_{D(\bar D)}(|{\bf p}|)
=\big(m_{D ({\bar D})}^2+|{\bf p}|^2\big)^{{1}/{2}}$, with 
$|{\bf p}|$, the magnitude of the momentum of the outgoing 
$D(\bar D)$ meson given by equation (\ref{pd}), and, 
$A_{\psi}(|{\bf p}|)$ is a polynomial in $|{\bf p}|$.
The parameter $\gamma_\psi$ is adjusted to reproduce the
vacuum decay widths of $\psi(3770)$ to $D^+D^-$ and
$D^0 \bar {D^0}$ \cite{amspmwg}.

We might note that the expressions of the decay widths
calculated using (I) the $^3P_0$ model as well as (II) a field theoretical
model of composite hadrons, as given by equations (\ref{dwpsi3770})
and (\ref{gammapsiddbar}) have the forms of a polynomial multiplied 
by a gaussian function of $|{\bf p}|$, which depends only 
on the masses of the charmonium and the $D(\bar D)$ mesons, 
as can be seen from equation (\ref{pd}). Within the 
$^3P_0$ model, the polynomial dependence 
was observed to lead to an initial increase 
followed by a drop and even vanishing of the decay width
with increase in the baryon density, in spite of the drop
in the masses of the outgoing $D$ and $\bar D$ mesons
\cite{friman,amarvepja}. A similar behaviour of an initial
increase of the decay width of $\psi(3770)$ to the neutral
$D\bar D$ pair with increase in the magnetic field
followed by a drop is observed in the magnetized 
hot nuclear (hyperonic) matter in the present work
for both the models.
As has already been mentioned, the masses of the $D$, $\bar D$ 
and the charmonium state, $\psi(3770)$ are the in-medium masses 
in the magnetized hot asymmetric nuclear (hyperonic) matter 
calculated in the chiral effective model. These are computed
including the effects of the Dirac sea of baryons and
due to PV mixing with additional contributions from 
lowest Landau levels for the charged open charm mesons
\cite{HQ_DW_DS_AM_SPM_2023}. 

Including the PV mixing effect, the expression
for the decay width is modified to 
\begin{eqnarray}
&&\Gamma^{[^3P_0]([FT])}_{PV}(\psi(3770) ({\bf 0}) \rightarrow D({\bf p}) {\bar D} (-{\bf p}))
=\frac{2}{3} \Gamma ^{[^3P_0] ([FT])}(\psi(3770) \rightarrow  D({\bf p}) 
{\bar D} (-{\bf p})) \nonumber \\
&+&\frac{1}{3} \Gamma ^{[^3P_0] ([FT])}(\psi(3770) \rightarrow D({\bf p}) {\bar D} (-{\bf p}))
(|{\bf p}|\rightarrow 
|{\bf p}|(M_{\psi}=M_{\psi}^{PV})),
\label{gammapsiddbar_mix}
\end{eqnarray}
where, $\Gamma(\psi(3770) \rightarrow  D({\bf p}) {\bar D}(-{\bf p})$ 
is given by expressions (\ref{dwpsi3770}) and (\ref{gammapsiddbar})
for computation using (I) the $^3P_0$ model and (II) the field
theoretical (FT) model of composite hadrons.
In equation (\ref{gammapsiddbar_mix}), the first term 
corresponds to the transverse
polarizations for the charmonium state, $\psi(3770)$
whose masses remain unaffected by the PV mixing,
whereas, the second term corresponds to the longitudinal component,
whose mass is modified due to the mixing with the pseudoscalar meson 
in the presence of the magnetic field.

\subsection{Production Cross-sections of Charmonium state $\psi(3770)$}

The production cross-section of vector meson, V, due to
scattering of particles $a_i$ and $b_i$ is given as
\cite{charm_prod_NM_B,am_spm_strange_spectral_fn,Elena_16,Elena_17,Elena_13_1,BW_CS_Haglin,BW_CS_Li}
\begin{equation}
\sigma _i (M)=6 \pi^2 
 \frac{{\Gamma_V^i}^*}{q({m^{*}_V},m^*_{a_i},m^*_{b_i})^2}
A_V (M),
\label{sigmaV_i}
\end{equation}
where,
\begin{equation}
A_V (M)=C \cdot \frac{2}{\pi} 
\frac {M^2 \Gamma_V^*}{(M^2-{m_V^*}^2)^2+(M\Gamma_V^*)^2}
\label{Spectral_V}
\end{equation}
is the relativistic Breit Wigner spectral function
expressed in terms of the invariant mass $M$, 
the mass and decay width of the vector meson,
$m_V^*$ and $\Gamma_V^*={\sum_j} {\Gamma_V^j}^* $,
in the magnetized matter, and,
$m^*_{a_i}$ and $m^*_{b_i}$ are the in-medium masses
of the scattering particles $a_i$ and $b_i$ in channel $i$.
In equation (\ref{sigmaV_i}), $q(m_V^*,m_{a_i}^*,m_{b_i}^*)$
is the momentum of the scattering particle $a_i(b_i)$
in the center of mass frame of the vector meson, $V$.
The normalization constant, $C$ in the 
spectral function is determined from
$\int _0 ^{\infty} A_V (M) d M=1.$
In the present work, the production cross-section
of the $\psi(3770)$ in channel $i=I,II$
arises from the two-body scatterings of 
(I) $D^+D^-$ and (II) $D^0 \bar {D^0}$ mesons respectively.
In the magnetized hot isospin asymmetric strange hadronic matter, 
these are calculated 
accounting for the Dirac sea and PV ($\psi(3770)-\eta_c'$,
$D-D^* (D^+-{D^*}^+,\; D^0-{D^*}^0)$ 
and $\bar D-{{\bar D}^*}
({D^-}-{{D^*}^-},\;{\bar {D^0}}-{{{\bar {D^*}}}^0})$)
mixing effects for the masses of these mesons. 
The decay width of $\psi(3770)\rightarrow D\bar D$ 
is obtained using the light quark antiquark pair creation models,
(I) $^3P_0$ model, as well as (II) a field theoretical
model of composite hadrons, described in the previous subsection.

The PV mixing effect introduces mass difference  
between the longitudinal and transverse components
of the vector charmonium state $\psi(3770)$
since the longitudinal component undergoes mass modification
due to mixing with the pseudoscalar meson $\eta_c'$,
whereas the transverse component is unaffected due to PV mixing.
In the presence of PV mixing, the production cross-section
of the vector meson, V, is given as 
\begin{equation}
\sigma_i (M)=6 \pi^2 
 \Bigg (\frac{1}{3}
\frac{{{\Gamma_V^*}^i}^{L}}{q({{m^*_V}^L},m^*_{a_i},m^*_{b_i})^2}
 A_V^L (M) +\frac{2}{3}
\frac{{{\Gamma_V^*}^i}^{T}}{q({{m^*_V}^T},m^*_{a_i},m^*_{b_i})^2}
 A_V^T (M) \Bigg),
\label{sigmaV_i_PV}
\end{equation}
for channel $i=I,II$ corresponding to the scattering of the 
charged $D^+D^-$ and neutral $D^0 {\bar {D^0}}$ mesons
respectively. In the above equation,
${m^*_V}^{T(L)}$ is the mass of the transverse (longitudinal)
component of the vector meson. ${{\Gamma_V^*}^i}^{T(L)}$ is
the decay width of the transverse 
(longitudinal) component of the vector meson
in channel $i$, with the total contribution to the decay
width of the transverse (longitudinal) component arising from
both the channels as given by
${\Gamma_V^*}^{T(L)}={{\Gamma_V^*}^I}^{T(L)}
+{{\Gamma_V^*}^{II}}^{T(L)}$.
In equation (\ref{sigmaV_i_PV}),
$A_V^{T(L)} (M)$ is the contribution from the transverse (longitudinal)
component to the spectral function of the vector meson, and is given as,
\begin{equation}
A_V^{T(L)} (M)=C \cdot \frac{2}{\pi} 
\frac {M^2 \Gamma_V^{*T(L)}}{\Big(M^2-{{m^{*T(L)}_V}^2\Big)}^2
+\Big(M\Gamma_V^{*T(L)}\Big)^2},
\label{Spectral_V_TL}
\end{equation}
with the constant $C$ determined from the normalization condition
\begin{equation}
\int _0 ^{\infty} \Big(\frac{1}{3} A_V^L (M)+ \frac{2}{3} A_V^T (M) 
\Big) d M=1.
\end{equation}
 
\section{Results and discussions}
\label{Sec.IV}
In the present investigation, we study the charmonium production
in magnetized hot isospin asymmetric strange hadronic 
matter. 
Strong magnetic fields are estimated to be produced
in ultra-relativistic peripheral collisions where the created 
matter is extremely diltute. In the present study, we study
the production cross-section of $\psi(3770)$, which is the lowest 
charmonium state which decays to $D\bar D$ in vacuum.
As has been described in the previous section, 
the mass modifications of the open charm ($D$ and $\bar D$) mesons 
and the charmoium state, $\psi(3770)$ are calculated 
within a generalized chiral effective model
\cite{amarindam,amarvdmesonTprc,amarvepja}.
The medium modifications of the masses
of these mesons are calculated including the effects of the 
baryonic Dirac sea (DS). In the presence of 
an external magnetic field, the masses of these mesons are 
additionally modified due to the mixing of the pseudoscalar
meson with the longitudinal component of the vector meson 
(PV mixing) and these modifications are observed to be quite
significant for high magnetic fields.  
The decay width of $\psi(3770)\rightarrow D\bar D$ 
in hot isospin asymmetric 
strange hadronic matter in the presence of an external magnetic field
is calculated from the medium modifications of the masses of 
the initial and final state mesons. 
Using the in-medium mass and decay width, the production cross-sections
of $\psi(3770)$, 
arising from scatterings of $D^+D^-$ as well as $D^0 \bar {D^0}$ mesons, 
are computed from the relativistic Breit-Wigner spectral
function. As can be seen later, these production cross-sections,
have distinct peak positions in the invariant mass spectra,
since the longitudinal component 
of $\psi(3770)$ undergoes a positive mass shift due to mixing
with $\eta_c(2S)$, whereas, the masses of the transverse components  
remain unaffected by PV mixing. 

The in-medium masses of the open charm mesons ($D$ and $\bar D$) 
within the chiral effective model are obtained, due to their
interactions with the baryons and the scalar mesons.
These are calculated by solving the dispersion
relation given by equation (\ref{dispersion}), where the self energies
of the $D$ and $\bar D$ are given by equations (\ref{selfd})
and (\ref{selfdbar}) respectively.
Within the mean field approximation,
the scalar (the nonstrange isoscalar, $\sigma$, 
the strange isoscalar $\zeta$ and 
the nonstrange isovector, $\delta$) fields, 
along with the dilaton field, $\chi$ and the vector fields,  
are solved from their coupled equations of motion, 
given by equations (\ref{sigma})--(\ref{phi}), 
for given  values of the temperature and the magnetic field 
for zero density, as well as, for hadronic matter for given
baryon density, strangeness, $f_s$ and isospin asymmetry parameter, 
$\eta$.
The charmonium mass shift is obtained from the value of the dilaton
field using equation (\ref{mass_shift_psi3770}).
In the present work, the anomalous magnetic moments (AMMs) 
of the baryons are taken into account.

Figures \ref{sg_zeta_dlt_eta0_rhb0_zero_density} and \ref{sg_zeta_dlt_eta3_rhb0_zero_density} 
show the dependence of the scalar fields $\sigma$, $\zeta$
and $\delta$
(which are related to the 
light quark condensates through equation (\ref{qqbar}))
on the magnetic field,
for $\rho_B=\rho_0$,
for symmetric and asymmetric (with $\eta$=0.3) matter respectively.
These are plotted for the cases of nuclear matter ($f_s$=0)
and for strange hadronic matter (with $f_s$=0.3) 
and compared to the results with zero baryon density. 
In the present work, the finite anomalous magnetic moments (AMMs)
of the baryons are incorporated. 
The values of the scalar fields 
are plotted with effects of the baryon Dirac sea 
and compared to the case when the Dirac sea effects are not considered
(shown as closely and widely spaced dotted lines for 
$\rho_B = 0$ and $\rho_0$, respectively).
The effects of the strangeness in the medium is observed
to lead to significant modifications to the scalar fields
both for symmetric and asymmmetric matter.
Contrary to the case of $\rho_B$=0 \cite{Sourodeep2023},
for the magnetized nuclear matter ($f_s$=0), 
at $\rho_B=\rho_0$, 
there is observed to be
a decrease in the magnitude of the scalar fields $\sigma$
($\sim m_u\langle  \bar u u\rangle +m_d \langle \bar d d\rangle$) 
as well as $\zeta$ ($\sim m_s \langle \bar s s\rangle$) 
fields with increase in the magnetic field, 
an effect called the inverse magnetic catalysis,
when the AMMs of the nucleons are taken into account.
As has been shown in Ref. \cite{Sourodeep2023}, 
the opposite effect 
of magnetic catalysis is observed in magnetized nuclear ($f_s$=0) matter
for $\rho_B=\rho_0$ at zero temperature, 
when the nucleon AMMs are ignored. In the magnetized nuclear matter,
for $\rho_B=\rho_0$, including the contributions of the 
Dirac sea and anomalous magnetic moments of the nucleons
leads to the inverse magnetic catalysis effect, also
at finite temperatures.
For magnetic field $eB = 4m_\pi^2$, the values of scalar field $\sigma 
 \left(\zeta \right)$ are observed to be $-58.39$  ($-96.89$), $-60.33$ 
($-97.32$), 
$-63.11$ ($-97.96$) and $-64.38$ ($-98.26$) MeV at temperatures $T = 0, 50, 
100$ and $150$ MeV, respectively. When the strength of magnetic field 
is increased to $eB=8 m_\pi^2$, these are modified to
$-49.04(-95.02)$, $-51.97(-95.57)$, $-56.42(-96.47)$ and $-40.44(-93.59)$ MeV,
respectively. 
We observe that the percentage drop in the magnitude of scalar 
field $\sigma$ ($\zeta$) is
$16.01\%(1.93\%)$, $13.86\%(1.80\%)$, $10.60\%(1.53\%)$
and $37.19\%(4.75$\%)  at T=0, 50, 100 and 150 MeV respectively. 
As can be seen, initially when temperature is increased from T =0  to 
100 MeV, the value of the percentage drop decreases. 
However, at large value of temperature ($T=150$ MeV) 
and stronger magnetic field, the magnitudes of the  scalar
fields decrease much more rapidly with increasing magnetic field. 
This implies that for large value of temperature and stronger magnetic 
field, the inverse magnetic catalysis is observed in nuclear matter.
 This observation is also consistent with lattice QCD observations where the
 finite temperature and stronger magnetic field is observed to show
the effect of inverse magnetic catalysis.
 
At zero temperature and for $\rho_B = \rho_0$, an increase in the value of isospin asymmetry 
parameter $\eta$
from zero to finite value (say $\eta$ = 0.3)   is observed to cause less drop  in the magnitude of scalar field. As temperature is increased the impact of $\eta$ is observed to weaken, i.e., values of scalar fields at finite $\eta$  start approaching the values observed at $\eta = 0$. In nuclear
 medium, at sufficiently high $T$ (for example at T = 150 MeV) the drop in the value of scalar fields is observed to be more at $\eta = 0.3$ as compared to $\eta = 0$.
 To quote in terms of numbers, we observe that for magnetic field  $eB = 8 m_\pi^2$ and isospin asymmetry $\eta = 0.3$, the values of scalar fields $\sigma (\zeta)$ are observed to be $-52.02 (-95.58)$, $-54.61(-96.09)$, $-59.03(-97.03)$
and $-37.96(-93.24)$ MeV at temperature $T = 0, 50, 100$ and $150$ MeV. 
As can be seen, at $T$=150 MeV, the magnitude of scalar fields 
at $\eta = 0.3$ are observed to be smaller as compared 
to the values quoted earlier for $\eta = 0$ case.
This implies that at high temperatures and for strong magnetic fields, 
inverse magnetic catalysis is more favored in asymmetric nuclear 
matter as compared to symmetric nuclear matter, although the effect
is small.
It may be noted that the anomalous magnetic moments of nucleons 
are observed
to affect the behavior of scalar fields more significantly 
in presence of finite Dirac sea as compared to when Dirac 
sea contributions are ignored. 
In the absence of the Dirac sea contributions, but accounting for
the finite anomalous magnetic moments of the baryons, 
the magnitude of scalar field increases with increasing the 
strength of magnetic field, at low and moderate temperature 
which implies that the magnetic catalysis effect is observed.
However, at high temperature such $T =  150$ MeV, the magnitude 
of scalar fields is observed to decrease with increase in the 
strength of magnetic field. 
It is observed that the impact of magnetic 
field is much more appreciable in presence of finite Dirac 
sea contributions, and the effect is particularly
significant for stronger magnetic field and high temperature case. 

We next study the effects of finite strangeness fraction on the
behaviour of the scalar fields ($\sigma$, $\zeta$ and $\delta$). 
At $\rho_B = \rho_0$, the inclusion of hyperons along with nucleons significantly 
alters the behaviour of scalar fields as a function of 
magnetic field, when the contributions due to the baryonic Dirac 
sea is taken into account.
The solutions of the scalar fields exist upto $eB\sim 5 m_\pi^2$,
in the strange hadronic matter ($f_s$=0.3) when the Dirac sea 
contributions and the AMMs of the baryons  are taken into account.
For $\rho_B=0$, the scalar fields, which have solutions for values
of $eB$ upto 4(2.8)$m_\pi^2$, when the Dirac sea contributions
of the nucleons (and hyperons) are taken into account,
along with the results when the DS effects are not considered,
are plotted in  figures \ref{sg_zeta_dlt_eta0_rhb0_zero_density} 
and \ref{sg_zeta_dlt_eta3_rhb0_zero_density}.
The non-existence of the solutions above a critical value of
the magnetic field can be understood in the following way.
For vacuum ($\rho_B=0$, $T$=0) of nuclear matter, 
the total scalar densities of the nucleons are solely 
due to contributions of the Dirac sea, 
and, the effective mass, $m_i^*$ satisfies the equation 
\begin{equation}
m_i^*-m_i= {C_i} \Bigg [ \frac{(q_i B)^2}{3 m_i^*}
+\{(\kappa_i B)^2 m_i^* +(|q_i|B)(\kappa_i B)\} 
\Big ( \frac{1}{2} +2 {\rm ln} \Big (\frac{m_i^*}{m_i}\Big)
\Big) \Bigg].
\label{meff_i_self_energy}
\end{equation}
where $C_i=A_i/(4\pi^2)$, with $A_i$ defined 
in equation (\ref{self_energy_i_DS}).
In the simplifying assumption of neglecting the logarithm term
in the last term on the right hand side of the above equation,
$m_i^*$ satisfies the quadratic equation 
\begin{equation}
(1-C_i(\kappa_i B)^2){m_i^*}^2
-\Big (m_i+\frac{C_i|q_iB|(\kappa_iB)}{2}\Big) m_i^*
-\frac{C_i (q_iB)^2}{3}=0.
\end{equation}
For real solutions for $m_i^*$, we should have 
$\Big (m_i+\frac{C_i|q_iB|(\kappa_iB)}{2}\Big)^2 \ge
\frac{4}{3} (1-C_i(\kappa_i B)^2)C_i (q_iB)^2$.
Also, for $C_i(\kappa_i B)^2 \ge 1$, i.e., for the value of the
magnetic field above a critical value
of $B_{crit}\sim {1/(C_i {\kappa_i}^2)}^{1/2}$, 
the above equation does not have
a positive solution for $m_i^*$,
both for the charged  ($q_i\ne 0$) and neutral ($q_i=0$) baryons.
It might be noted that the above equation 
corresponds to the case when there are no meson-meson 
interaction terms for the scalar fields in
the Lagrangian. 
In the presence of these quartic interaction terms 
(as given by equation (\ref{L_0})) in the chiral model, 
the effective masses of the baryons, 
$m_{i}^{*}(=-g_{\sigma i}\sigma-g_{\zeta i}\zeta 
-g_{\delta i} \delta)$,
are obtained from the scalar fields,
which are solved, along with the dilaton field, $\chi$
and the vector fields, from their coupled equations 
of motion given by equations (\ref{sigma}) -- (\ref{phi}).
In the absence of hyperons,
the scalar densities of nucleons, for $\rho_B=0$, are solely
due to the negative contributions of the Dirac sea,
which leads to an increase in the magnitudes of the scalar fields
(hence an enhancement of the light quark condensates)
with increase in the strength of the magnetic field.
This effect of magnetic catalysis for $\rho_B=0$ is observed, 
for both the cases of without and with the effects 
from the nucleon anomalous magnetic moments (AMMs)
\cite{haber2014,Arghya2018,Sourodeep2023,Parui_D12023,HQ_DW_DS_AM_SPM_2023}.
In the present investigation of strange hadronic matter
at finite temperatures in the presence of a magnetic field,
for $f_s$=0.3 and $\rho_B=\rho_0$, the number densities for the hyperons
remain negligible (except for $\Xi^-$) and hence the contributions 
to the total scalar densities of these heyperons turn out
to be negative, solely due to the contributions from the 
Dirac sea, as given by equation (\ref{rhosi_DS}). It is 
observed that the solutions for the scalar fields do not
exist for values of the magnetic field above a critical 
value ($eB_{crit}\sim 5 m_\pi^2$ for $f_s=0.3$ and $\eta$=0.3),
similar to the situation of the vacuum in nuclear matter,
for reasons described above.
%
%
Also, as the magnetic field is raised, for the hyperons,
which are absent for $\rho_B=\rho_0$, the total scalar
densities are negative, solely due to  
the Dirac sea contributions,
leading to an increase in the magnitudes of scalar fields 
($\sigma$ and $\zeta$), leading to magnetic catalysis effect 
in the magnetized strange hadronic matter (with $f_s=0.3$),
contrary to the opposite behaviour of the scalar fields 
(hence of the light quark condensate)
with magnetic field for the nuclear matter 
at $\rho_B=\rho_0$, when the AMMs of the baryons 
are taken into account.
In the presence of Dirac sea contributions
from the nucleons (hyperons along with nucleons), 
the effect of magnetic catalysis (an increase in the magnitude 
of scalar fields with increasing  magnetic field strength) is 
observed for $\rho_B=0$, as might be seen from figures
\ref{sg_zeta_dlt_eta0_rhb0_zero_density} and 
\ref{sg_zeta_dlt_eta3_rhb0_zero_density}.

In figures \ref{md_eta3_fs0_Temp_rhb0_zero_density} and \ref{md_eta3_fs3_Temp_rhb0_zero_density},
the masses of the $D (D^0,D^+)$ mesons are plotted as functions
of $eB/m_\pi^2$ for the magnetized asymmetric matter 
(with $\eta$=0.3) at nuclear matter saturation density ($\rho_B=\rho_0$) 
for the nuclear ($f_s$=0) and hyperonic matter (with $f_s$=0.3)
respectively.
In the left panel of these figures, 
results for the mass modifications of $D$ mesons are also shown at
$\rho_B = 0$, considering magnetized Dirac sea of nucleons
and in the right panel, due to both the nucleons as well as hyperons.
The anomalous magnetic moments (AMMs)
of the baryons are taken into account in the present study.
The $D^+$ and $D^0$ masses are shown for values of the temperature 
T=50, 100 and 150 MeV. 
The in-medium behaviors of the scalar fields 
are reflected in the medium modification of $D$ meson properties.
The lowest Landau level (LLL) contribution is taken into account
for the charged $D^+$ meson.
At finite baryon density,
in the absence of Dirac sea contributions, when the  PV mixing 
is not taken into account, one observes  the mass
of the neutral $D$ meson to be insensitive to the variation 
in the magnetic field whereas $D^+$ mass shows a steady increase
due to the LLL contributions. 
Including the Dirac sea contributions and the anomalous magnetic 
moments of nucleons, initially the in-medium mass of $D^{+}$ meson 
is seen to increase with magnetic field upto a certain value and  
then starts decreasing with further increase in $eB$. 
The drop in the mass at higher values of magnetic field is observed
to be sharper as the temperature is raised from T=50 MeV
to T=100 and 150 MeV. In the presence of 
PV ($D-D^*$) mixing, there is seen to be a drop in the mass
of both the $D^+$ and $D^0$ mesons, which is observed to be much
more pronounced for the neutral $D$ meson. 
The effect of the  isospin asymmetry is observed to be small.
For example, for $\rho_B=\rho_0$ and $f_s=0$ the value of the
$D^+ (D^0)$ mass (in MeV) is observed to be
1736.5 (1585.9) for $eB=10 m_\pi^2$ for T=50 MeV
and 1615.86 (1416.3) at T=150 MeV for the same value 
of the magnetic field for $\eta$=0.3, when the Dirac sea 
as well as PV mixing are taken into account, in addition to 
LLL contribution for $D^+$ meson, which may be compared to
the $\eta$=0 values of 1732.85 (1554.5) and
1625.6 (1414.9) at T=50  and T=150 MeV respectively.
Thus the isospin asymmetry effect is small, but larger for 
neutral $D$ meson at smaller temperature of T=50 MeV,
whereas, for $D^+$ mass, the effect of isospin asymmetry is
observed to be marginal.

The effect of strangeness is shown in figure \ref{md_eta3_fs3_Temp_rhb0_zero_density}, 
which shows the magnetic field dependence of the $D$ meson masses 
for the asymmetric ($\eta = 0.3$) hyperonic matter (with $f_s$=0.3)
for $\rho_B=\rho_0$. Similar to the case of asymmetric
(with $\eta$=0.3) nuclear matter ($f_s$=0)
as shown in figure \ref{md_eta3_fs0_Temp_rhb0_zero_density}, 
in the absence of the Dirac sea contributions, the $D^0$ mass is
observed to be insensitive to the variation in the magnetic field,
whereas the LLL contribution leads to a monotonic increase 
in the mass of $D^{+}$ mesons with
$eB$, when the PV ($D-D^*$) mixing is not taken into account. 
The PV mixing leads to a drop in the masses of both $D^+$ and $D^0$,
with a much larger decrease for $D^0$ mass, as has been observed
for the case of magnetized nuclear matter ($f_s$=0). 
At $\rho_B = \rho_0$, in the presence of the Dirac sea contributions, the solutions
for the scalar fields exist only upto a value of $eB=5 m_\pi^2$,
as has already been mentioned. There is observed to be an increase
in the mass of $D^+$ as was observed for $f_s$=0 case 
upto $eB=5 m_\pi^2$. For zero magnetic field, 
when the strangeness fraction $f_s$ is changed from 0 to 0.3, 
the value of the mass of $D^+$ ($D^0$) is modified from 1797.9 
(1792.2) to  1791 (1799.8) for T=50 MeV, and 
1801.86 (1798.1) to 1794.8 (1803.9) for T=150 MeV.
For $eB=5 m_\pi^2$, the modifications in the masses due to increase
in $\eta$ from 0 to 0.3 are 1857 (1788.4) to 1855.7 (1801.8) 
and 1884.5 (1816.5) to 1882.1 (1826.1) for T=50 and T=150 MeV respectively.
The effects of the isospin asymmetry on the $D^+$ and $D^0$ 
meson masses for the strange hadronic matter 
are thus observed to be similar to the symmetric case.
As has already been mentioned, at zero baryon density,
in the presence of magnetized Dirac sea of nucleons (nucleons and hyperons),
the solutions of scalar fields exist upto $eB\sim 4 (2.8) m_\pi^2$.
The $D$ mesons masses for $\rho_B = 0$ with the Dirac sea contributions
are plotted upto these values of magnetic field 
in figures \ref{md_eta3_fs0_Temp_rhb0_zero_density}
and \ref{md_eta3_fs3_Temp_rhb0_zero_density}. 
The LLL contribution (for the charged $D^+$ meson) causes
an increase in the mass as a function of magnetic field 
both at zero baryon density as well as at $\rho_B=\rho_0$, 
whereas the mass decreases due to the PV
mixing. 
At zero baryon density, the effects of Dirac sea contributions
due to nucleons (nucleon and hyperons) on the in-medium masses
of the $D^+$ and $D^0$ mesons are observed to be marginal
as compared to when these effects are ignored, 
upto a value of $eB$ around $3.5(2) m_\pi^2$ 
beyond which there is observed to be a drop
with further increase in the magnetic field. 

The dependence of in-medium masses of $D^{-}$ and $\bar{D^{0}}$ mesons
on the magnetic field 
are shown in figures  \ref{mdbar_eta3_fs0_Temp_rhb0_zero_density} and \ref{mdbar_eta3_fs3_Temp_rhb0_zero_density}
for values of the isospin asymmetry parameter $\eta=0.3$, 
for magnetized nuclear ($f_s = 0$) and hyperonic (with $f_s$=0.3) matter  
respectively, at $\rho_B = \rho_0$, 
along with 
the results for zero baryon density.
At $\rho_B = \rho_0$, in magnetized nuclear matter ($f_s$=0), 
for a given magnetic field strength, the in-medium masses of $\bar{D}$ 
mesons are observed to be larger than the masses of the $D$ mesons.
However, the trend of variation of in-medium masses of $\bar{D}$ mesons 
as  functions of the magnetic field are found to be similar 
to those of the $D$ mesons. The effects of the isospin asymmetry 
of the nuclear medium on the masses of the $\bar D$ mesons
are observed to be small compared to
the effects from the Dirac sea and PV mixing effects,
similar to what was observed for the $D$ mesons.
As has already been mentioned,
at finite strangeness fraction with $f_s = 0.3$  and $\rho_B = \rho_0$,
the solutions for the scalar fields exist only upto $eB=5 m_\pi^2$,
when the Dirac sea contributions are considered by summing over the tadpole
diagrams in the weak magnetic field limit. 
Similar to the mass of the $D$ meson, the $\bar{D}$ meson masses
are observed to increase with magnetic field upto $5m_\pi^2$.
As has been observed for the $D$ mesons, the masses of $\bar D$ mesons
have very small effect from the isospin asymmetry for the nuclear
as well as hyperonic matter. 
For $\rho_B = 0$, the behaviour of $\bar{D}$ masses
are observed to be similar to the masses of the $D$ mesons.

At finite density, the dominant magnetic field effects on the $D$ and $\bar D$ mesons
are due to the Dirac sea and PV mixng and the effects of isospin
asymmetry are observed to be marginal.
The qualitative trends of the $D$ and $\bar D$ masses both 
for the nuclear and
hyperonic matter are observed to be similar at temperatures
(T=50, 100 and 150 MeV) considered in the present study.
However, in nuclear matter (both for $\eta$=0 and 0.3), 
the masses of all these mesons are observed to initially
increase followed by a drop, but the value of $eB$ where the masses
start decreasing, are observed to be smaller when the
temperature is raised and the value is much smaller
for T=150 MeV.

The in-medium mass of charmonium state $\psi(3770)$ 
as function of $eB$ is shown in figure 
\ref{psi3770_eta3_Temp_rhb0_zero_density} 
for the asymmetric (with $\eta$=0.3)
nuclear matter ($f_s$=0) and the hyperonic matter (with $f_s$=0.3), 
at $\rho_B = \rho_0$, 
along with results for
$\rho_B = 0$.
At finite $\rho_B$, in the presence of Dirac sea contributions,
in nuclear matter, both for symmetric and asymmetric cases,
there is observed to be an increase in the mass followed by a drop
with further increase in the magnetic field. However, similar
to the cases of $D$ and $\bar D$ meson masses, the effects
of isospin asymmetry are observed to be small for the
charmonium ($\psi(3770)$) mass.  The value of
$eB$ for which the behaviour changes from an increase in mass to a 
drop, is observed to be smaller when the temperature is raised. 
As has been already mentioned, the masses of the charmonium
states are obtained from the medium change of the gluon 
condensate and also depend on the wave function of the particular
state considered, using equation (\ref{mass_shift_psi3770}).
The contribution from PV mixing leads to a negative (positive)
contribution to the mass of the pseudoscalar (longitudinal 
component of the vector) meson. This is calculated from 
the phenomenological Lagrangian corresponding to the process
$V\rightarrow P\gamma$. For T=150 MeV, the mass of the 
$\psi(3770)\equiv \psi(1D)$ turns out to be smaller
than the mass of $\eta_c'\equiv \eta_c(2S)$ for $eB$ larger
than around 8.5 $m_\pi^2$, due to which the process 
$\psi(3770)\rightarrow \eta_c'\gamma$ is no longer kinematically 
possible. The effect due to PV mixing on the mass of $\psi(3770)$
exist upto $eB=8.5 m_\pi^2$ for T=150 MeV, for symmetric and asymmetric
nuclear matter. These can be seen from panel
(e) of figure \ref{psi3770_eta3_Temp_rhb0_zero_density} for asymmetric nuclear
matter. At $\rho_B = \rho_0$, in the presence of strangeness fraction, including
the Dirac sea contributions, lead to solutions
for the mass of $\psi(3770)$ upto $eB = 5 m_\pi^2$,
which is observed to increase with the magnetic field
for the temperatures T=50, 100 and 150 MeV, both for isospin
symmetric and asymmetric cases. This can be observed
from panels (b), (d) and (f) for asymmetric hyperonic matter
in figure \ref{psi3770_eta3_Temp_rhb0_zero_density}.
At zero baryon density, with magnetized Dirac sea 
of nucleons (and hyperons), the effective masses of charmonium
state  (calculated from the medium change of the dilaton field)
are obtained upto $4 (2.8) m_\pi^2$, due to non-existence 
of solutions of scalar fields above these values of magnetic field. 
The DS effects are observed to be marginal for $eB$ 
upto around $3.5 (2)m_\pi^2$ above which there is observed to be
a drop in the charmonium mass.

Figure \ref{Decaywidth_psi3770_3p0_eta3_Temp_rhb0_zero_density} shows 
the dependence of the partial decay width 
of $\psi(3770)\rightarrow D\bar D$, along with the 
contributions from the subchannels 
(I) $\psi(3770)\rightarrow D^+D^-$ and
(II) $\psi(3770)\rightarrow D^0 \bar {D^0}$, 
computed using the $^3P_0$ model.
These decay widths are shown
for magnetized matter for $\rho_B = 0$
and also for 
asymmetric (with $\eta$=0.3) nuclear
as well as hyperonic matter (with $f_s$=0.3)
at $\rho_B=\rho_0$ and for values of temperature
as 50, 100 and 150 MeV.
The medium modifications of the decay widths
arise due to the medium modifications of the masses
of the intial and final states as calculated using the chiral model
including the effects of the baryonic Dirac sea, with additional
contributions from PV mixing as well as the lowest Landau level 
(LLL) for the charged open charm mesons. The effects due to the 
PV mixing ($\psi(3770)-\eta_c'$, $D-D^*$ and $\bar D-\bar {D^*}$)
are considered. These decay widths are compared with the values 
when the Dirac sea effects are not considered.
When the PV mixing effects are not taken into account,
the decay width of $\psi(3770)$ to $D^0\bar {D^0}$
remains insensitive to the change in the
magnetic field when the Dirac sea contributions are not
considered as the masses are almost unaffected by the
magnetic field for the neutral open charm mesons 
\cite{Reddy2018} and the charmonium state \cite{Amal2018}.  
However, in the absence of Dirac sea effects, the decay width to
the charged $D\bar D$ final state is observed to decrease 
with increase in the magnetic field and vanish as the magnetic
field is further increased, when the PV mixing effects
are still not considered \cite{charmdecay_mag}. 
This is due to LLL contributions
which leads to a rise in the masses of the charged open 
charm mesons \cite{Reddy2018}. 
The incorporation of the PV mixing effects
(due to $\psi(3770)-\eta_c'$, $D(\bar D)-D^*(\bar {D^*})$) 
is observed to lead to
significant modifications to the decay widths
of $\psi(3770)$ to the charged and neutral $D\bar D$ 
due to the magnetic field, 
both for nuclear and strange hadronic matter.
Accounting for the PV mixing, but when the Dirac sea (DS)
effects of the baryons are not taken into consideration,
within the $^3P_0$ model, there is observed to be 
an intial slow decrease in the decay width for the 
charged $D\bar D$ mesons channel with the increase 
in the magnetic field, which remains almost constant
at higher values of the magnetic field,
whereas, the neutral $D\bar D$ channel shows an initial 
increase followed by a drop with further increase
in the value of the magnetic field.
This behaviour is observed for symmetric as well as
asymmetric (with $\eta$=0.3) for both nuclear matter
and hyperonic matter. This can be seen from figure 
\ref{Decaywidth_psi3770_3p0_eta3_Temp_rhb0_zero_density}, 
plotted for asymmetric nuclear and hyperonic matter.
In the presence of Dirac sea contributions,
the behaviour of the decay widths with magnetic field
remain similar and the effects from PV mixing are
observed to dominate over the effects of Dirac sea.
The decay widths, obtained upto $eB\sim 4(2.8)m_\pi^2$ 
with the effects of Dirac sea of the nucleons (and hyperons)
for $\rho_B=0$, have similar trend as for $\rho_B=\rho_0$. 

The magnetic field dependence of the charmonium decay widths 
are shown for asymmetric nuclear and hyperonic matter 
in figure \ref{Decaywidth_psi3770_FT_eta3_Temp_rhb0_zero_density} using
the field theoretical (FT) model of composite hadrons.
At $\rho_B = \rho_0$, the decay width in asymmetric nuclear matter
in the charged $D\bar D$ channel
is observed to decrease with the magnetic field 
upto $eB\sim 6 m_\pi^2$,
beyond which there is observed to be a slow increase. 
The values of the decay width (in MeV) 
are observed to be 31.48, 19.12 and 24.86 for values
of $eB$ (in units of $m_\pi^2$) as 0, 6 and 10 respectively
for T=50 MeV when the DS effects are taken into account.
On the other hand, the decay width for the neutral $D\bar D$
channel is observed to have a substantial increase with the
magnetic field, from the value of 22.68 MeV at $eB=0$
reaching a maximum of around 78.3 MeV for $eB\sim 7.5 m_\pi^2$, 
for the same temperature,
beyond which there is observed to be a slow drop as the magnetic field
is further increased. This leads to the maximum value
of the total decay width to be 98.9 MeV for 
$eB=8.5 m_\pi^2$ for T=50 MeV in asymmetric nuclear matter
when the DS effects are taken into account.
Similar behaviours of the decay widths are observed
for the temperatures T=100 and 150 MeV for asymmetric nuclear
matter.
The Dirac sea (DS) contributions are observed to lead 
to smaller values of decay width, although the changes due
to DS effects are marginal.
In the presence of hyperons in the medium, the decay widths
follow similar trend as for the nuclear matter for T=50, 100
and 150 MeV, as can be observed 
from figure \ref{Decaywidth_psi3770_FT_eta3_Temp_rhb0_zero_density}.
At $\rho_B = \rho_0$, the decay widths are plotted only upto $eB=5 m_\pi^2$ 
for strange hadronic matter, when the DS effects are 
taken into account since the solutions for the scalar fields
do not exist for $eB$ larger than $5 m_\pi^2$, as has already 
been mentioned. 
In figure \ref{Decaywidth_psi3770_FT_eta3_Temp_rhb0_zero_density}, 
the decay widths of $\psi(3770)$ to charged and neutral $D\bar D$
mesons, along with the total of both subchannels,
obtained using the field theoretical model,
are also plotted for $\rho_B = 0$, and, are
observed to have similar trend as obtained using the $^3P_0$ model.
The magnetic field effect on the decay width 
due to PV mixing is observed to dominate over the effect 
from Dirac sea within the $^3P_0$ model as well as the 
field theoretical (FT) model of composite hadrons.

The production cross-sections of $\psi(3770)$, arising from
the scattering of (I) $D^+D^-$ and (II) $D^0\bar {D^0}$ mesons
are plotted as functions of the invariant mass
for magnetized asymmetric (with $\eta$=0.3) nuclear ($f_s$=0) 
and hyperonic (with $f_s$=0.3) matter 
with $\psi(3770)\rightarrow D^+D^- (D^0 \bar {D^0})$
decay width computed using the $^3P_0$ model
in figures
 \ref{REV2_Cross_section_3p0_psi3770_eta3_fs0_2_5_8mpi2}
and  \ref{REV2_Cross_section_3p0_psi3770_eta3_fs3_2_5_8mpi2}
and using FT model, in figures 
\ref{REV2_Cross_section_FT_psi3770_eta3_fs0_2_5_8mpi2}
and \ref{REV2_Cross_section_FT_psi3770_eta3_fs3_2_5_8mpi2}, 
at $\rho_B = \rho_0$ for T=50, 100 and 150 MeV.
These are plotted for values of $eB = 2.5 m_\pi^2$ and $eB = 8 m_\pi^2$.
For $\rho_B = 0$, for which the results for the production cross-sections
with the Dirac sea effects due to the nucleons (and hyperons)
exist for $eB$ upto around $4(2.8)m_\pi^2$,
the production cross-sections are shown 
for $eB = 2.5 m_\pi^2$ (left panels of 
Figs. \ref{REV2_Cross_section_3p0_psi3770_eta3_fs0_2_5_8mpi2}
--\ref{REV2_Cross_section_FT_psi3770_eta3_fs3_2_5_8mpi2}). 
The PV ($\psi(3770)-\eta_c'$) mixing modifies the mass of the longitudinal 
component of the charmonium state, $\psi(3770)$, whereas, 
the transverse components remain unaffected due to PV mixing.
There is observed to be
well-separated distinct peak positions
in the production cross-sections of $\psi(3770)$ due to scattering 
of (I) $D^+D^-$ and (II) $D^0 \bar{D}^0$ plotted as functions 
of the invariant mass, for the higher value
of the magnetic field ($eB=8m_\pi^2$), as the PV mixing effects
are more dominant with increase in the magnetic field.
For magnetized asymmetric (with $\eta$=0.3) 
nuclear matter, at $\rho_B = \rho_0$, 
and for the values of temperatures
of T=50, 100 and 150 MeV respectively, the production cross-sections
due to the scattering of the $D^+D^-$ and $D^0 {\bar {D^0}}$
are plotted as functions of the invariant mass in panels (a), (c), and (e) 
for $eB= 2.5 m_\pi^2$ and in panels (b), (d) and (f) for $eB=8m_\pi^2$
in figure \ref{REV2_Cross_section_3p0_psi3770_eta3_fs0_2_5_8mpi2}.
For $D^{+}D^{-} (D^0 \bar {D^0})$ channel, these are observed 
to be maximum at positions (in MeV)
of 3706 (3706), 3722 (3722) and 3726 (3726) 
when the Dirac sea (DS) effects are taken into account, 
and 3708 (3708), 3724 (3722) and 3728 (3728),
when these are not taken into consideration.
It might be noted here that the production cross-section
due to scattering of $D^+D^-$ as well as $D^0 \bar {D^0}$
has contributions due to the transverse and longitudinal
components of $\psi(3770)$, which have different masses
due to PV mixing. The masses (in MeV) of the longitudinal (transverse)
component of $\psi(3770)$ for $\rho_B=\rho_0$, $f_s=0$
and $\eta=0.3$ and T=50, 100 and 150 MeV, are observed 
to be 3723.6 (3703.4),  3745.8 (3719.2) and  3744.1 (3725),
in the presence of DS effects, and, 3725 (3705), 3740 (3720.7) 
and 3725 (3751), when the DS effects are neglected.  
It is thus observed that the production cross-sections of
$\psi(3770)$ due to scattering of the charged as well as
neutral $D\bar D$ are peaked at position close to the 
transverse mass of $\psi(3770)$, for $eB=2.5m_\pi^2$.
At $\rho_B = \rho_0$, the production cross-sections of $\psi(3770)$ 
due to the scatterings
of (I) $D^+D^-$ and (II) $D^0 \bar {D^0}$ mesons are observed
to be very similar for the magnetic field $eB=2.5 m_\pi^2$,
with the value of the peak height to be marginally larger
for the channel (I) $D^+D^-$ as compared to (II) $D^0 \bar {D^0}$, 
as might be seen from  panels (a), (c) and (e) for T=50, 100 and 150 MeV.
However, both with as well as without DS effects,
for $\rho_B=0$ and $eB=2.5 m_\pi^2$, the peak height 
is observed to be much larger for the $D^0\bar {D^0}$ 
as compared to the value for $D^+D^-$ scattering. 
E.g., the value of the peak height is 82.5mb 
for $D^0\bar {D^0}$ is much marger as compared to the value of 
59.8 mb for $D^+D^-$ scattering, 
for T=50 MeV, in the presence of DS effects. 
The effect of temperature on the production cross-sections
is observed to be very small for $eB=2.5m_\pi^2$.
At the higher magnetic field of $eB=8 m_\pi^2$,
the masses with and without PV mixing are quite different,
as can be seen from figure \ref{psi3770_eta3_Temp_rhb0_zero_density},
with the values of 3644.4 (3774.26),
3682.9 (3806.65) and 3442 (3608.9) MeV for 
T=50, 100 and 150 MeV respectively, when the Dirac sea (DS)
effects are taken into account, and at 3708.25 (3828.1),
3721.6 (3839.1) and 3702.2 (3823) MeV when these are not taken into
consideration. 
The masses (in MeV) of the longitudinal (transverse)
component of $\psi(3770)$ for $\rho_B=\rho_0$, $f_s=0$,
$\eta=0.3$, $eB=8 m_\pi^2$, and T=50, 100 and 150 MeV, 
are observed to be 3774.3 (3644.4), 3806.7 (3682.9)
and 3608.9 (3442), in the presence of DS effects, 
and, 3828.1 (3708.2), 3839.1 (3721) and
3823 (3702), when the DS effects are neglected.  
For the higher value of the magnetic field, $eB=8 m_\pi^2$,
as can be seen from panels (b), (d) and (f) in figure 
\ref{REV2_Cross_section_3p0_psi3770_eta3_fs0_2_5_8mpi2}
for temperatures T=50, 100 and 150 MeV respectively,
the larger mass difference in the transverse and longitudinal
components of $\psi(3770)$ leads to the peaks to be well separated,
both for $\rho_B=\rho_0$ as well as for $\rho_B$=0.
The positions of the peaks are 
observed to be close to the longitudinal (transverse) 
mass of $\psi(3770)$, due to the scattering of
$D^{+}D^{-}$ ($D^{0}\bar{D}^0$) mesons.
The production cross-sections from the $D^+D^-$ scatteting is observed 
to have additional peaks at positions close to the transverse (lower) 
mass of $\psi(3770)$ for T=50 and 100 MeV for $\rho_B=\rho_0$ 
(with $\eta$=0.3, $f_s$=0), in the absence of the Dirac sea effects,
as can be seen from panels (b) and (d) (as the widely-spaced dotted lines) 
of Fig. \ref{REV2_Cross_section_3p0_psi3770_eta3_fs0_2_5_8mpi2}. 
It is observed that the individual production 
cross-sections are diminished for the higher value of the 
magnetic field, $eB=8 m_\pi^2$,
with the heights of the peaks (in mb) to be  given as
17.62 (17.57) and 26.12 (25.56)
for T=50 (100) MeV and $\rho_B = \rho_0$, 
due to channels (I) and (II) respectively, 
in the presence of DS effects, as compared
to the values of the heights to be 55.39 (54.85) and 
54.15 (52.91) for $eB=2.5 m_\pi^2$. At $\rho_B = \rho_0$, 
for T=150 MeV and $eB=8m_\pi^2$, 
with DS effects, the peak heights (in mb) are observed to be 
23.26 and  19.49 for channels (I) and (II) respectively,
which are similar to the values of 24.52 and  18.94  to the case 
when DS effects are not taken into account. 

In the presence of hyperons in the medium,
when the Dirac sea effects are taken into account, there is observed
to be an increase in the mass of charmonium state $\psi(3770)$
as calculated within the chiral effective model using equation
(\ref{mass_shift_psi3770}), whose longitudinal component has 
a further positive shift due to the PV mixing, for T=50, 100 
and 150 MeV, as can be seen 
for asymmetric ($\eta$=0.3) strange ($f_s$=0.3)
hadronic matter in figure \ref{psi3770_eta3_Temp_rhb0_zero_density}.
The behaviour of the charmonium mass is reflected
in the peak positions of the production cross-section
as plotted for the magnetized asymmetric strange hadronic
matter in figure \ref{REV2_Cross_section_3p0_psi3770_eta3_fs3_2_5_8mpi2}.
The production cross-sections of $\psi(3770)$ are plotted in figure
\ref{REV2_Cross_section_3p0_psi3770_eta3_fs3_2_5_8mpi2}
for asymmetric (with $\eta$=0.3) strange (with $f_s$=0.3)
hadronic matter for $eB=2.5 m_\pi^2$ and $eB=8 m_\pi^2$,
for T=50, 100 and 150 MeV and $\rho_B = \rho_0$ (also, at $\rho_B = 0$ for $eB = 2.5 m_\pi^2$).
For $eB=2.5 m_\pi^2$, the behaviour of the production cross-sections 
in the channels (I) and (II) are observed to be similar to the case of
asymmetric nuclear matter (shown in panels (a), (c) and (e)
in figure \ref{REV2_Cross_section_3p0_psi3770_eta3_fs0_2_5_8mpi2}).
However, the peak positions are shifted to higher values
when the DS effects are taken into consideration as compared
to when the Dirac sea of the baryons are ignored.
At $\rho_B = \rho_0$ and $eB=2.5 m_\pi^2$, the DS effects are observed to be larger
as compared to the magnetized nuclear matter shown in figure
\ref{REV2_Cross_section_3p0_psi3770_eta3_fs0_2_5_8mpi2}.
As has already been mentioned, at $\rho_B = \rho_0$, for $eB$ higher
than $5m_\pi^2$, the solutions for the scalar fields 
do not exist for the strange hadronic matter, when the 
Dirac sea effects of baryons are taken into account 
in the weak magnetic field limit as used for
the computation of the baryon self-energy accounting for
the magnetized Dirac sea \cite{Arghya2018}.
The dependence of the production cross-sections
on the invariant mass are observed to be similar 
to the case of magnetized nuclear matter for the higher
value of the magnetic field, $eB=8 m_\pi^2$, when the 
Dirac sea effects are not taken into account.

The production cross-sections of $\psi(3770)$
are plotted for magnetized asymmetric (with $\eta$=0.3) 
nuclear ($f_s$=0) and hyperonic
(with $f_s$=0.3) matter in figures 
\ref{REV2_Cross_section_FT_psi3770_eta3_fs0_2_5_8mpi2}
and \ref{REV2_Cross_section_FT_psi3770_eta3_fs3_2_5_8mpi2}
using FT model, for values of $eB=2.5 m_\pi^2$ and $eB=8 m_\pi^2$.
The larger values of the charmonium decay width 
obtained using the field theoretical (FT) model
as compared to the values calculated within the $^3P_0$
model are observed as a broadening of the peaks
in the invariant mass plot of the production cross-section of
$\psi(3770)$. 
The effects from the temperature and strangeness are observed 
to be significant in the present study of the production
cross-section of the charmonium state $\psi(3770)$
for higher values of the magnetic field,
which are obtained from the in-medium mass and decay widths
of the charmonium state in the magnetized nuclear (hyperonic)
matter. The PV mixing effect is observed to dominate over 
the Dirac sea effect.
 In the invariant mass plot of production 
cross-sections of $\psi(3770)$
the production cross-sections for the scatterings from
(I) $D^+D^-$ and (II) $D^0\bar {D^0}$ are observed to be peaked
at the positions close to the mass of the longitudinal 
and transverse components of the charmonium state $\psi(3770)$
for higher values of the magnetic field ($eB=8m_\pi^2$) 
for magnetized nuclear matter, when the DS effects are taken 
into consideration for $\rho_B=\rho_0$, as well as,
for $\rho_B$=0, when the Dirac sea effects are not taken into
account. In the absence of DS effects, for $eB=8m_\pi^2$,
both for the nuclear and hyperonic matter, additional peaks 
are observed for T=50 and 100 MeV for the production
cross-section of the charmonium state in the charged $D\bar D$
cahnnel, close to the transverse mass of $\psi(3770)$
at $\rho_B=\rho_0$, as can be seen from panels (b) and (d)
of figures \ref{REV2_Cross_section_3p0_psi3770_eta3_fs0_2_5_8mpi2}
--\ref{REV2_Cross_section_FT_psi3770_eta3_fs3_2_5_8mpi2}.
For T=150 MeV, the peaks are observed to be appreciably well-separated
as compared to the lower temperatures (T=50 and 100 MeV).
The production cross-sections of $\psi(3770)$ can have 
consequences on the dilepton spectra as well
as the production of the charmonium as well as open charm
mesons produced in ultra-peripheral ultra-relativistic
heavy ion collison experiments, since the magnetic field
created in these collisions are extremely large and the
charm mesons are created at the early stage of the collision.  

\section{Summary }
\label{Sec.V}
To summarize, in the present paper, we have investigated 
the charmonium production cross-sections
due to scatterings of $D^+D^-(D^0\bar {D^0})$ mesons,
in the presence of an external magnetic field at
finite temperatures, for $\rho_B=0$ as well as 
in isospin asymmetric nuclear (hyperonic) matter
for $\rho_B=\rho_0$.
In the peripheral ultra-relativistic heavy ion
collisions, strong magnetic fields are produced. However,
since the created matter has extremely low density, 
we investigate the production cross-sections of the charmonium state, 
$\psi(3770)$, which is the lowest state which decays to
$D\bar D$ in vacuum. 
The production cross-sections are calculated from 
the Breit-Wigner spectral function
of the charmonium state, which is expressed in terms of
its in-medium mass and decay width.
The effects of Dirac sea as well as PV mixing, in addition
to the lowest Landau level (LLL) contributions for the charged mesons,
are taken into consideration to calculate the
masses of the open charm ($D$ and $\bar D$) mesons and the charmonium
state $\psi(3770)$ and the subsequent effect on the partial
decay width of $\psi(3770)\rightarrow D\bar D$ in magnetized
isospin asymmetric nuclear (hyperonic) matter for $\rho_B=\rho_0$
as well as for $\rho_B=0$, at finite temperatures.
The decay widths are calculated using two light quark-antiquark
models: (I) the $^3P_0$ model and (II) a field theoretical (FT)
model of composite hadrons, and their effects on the production
cross-section of $\psi(3770)$ are studied in the present work.

Within a chiral effective model, the in-medium masses of the 
open charm mesons are calculated from their interactions 
with the scalar (isoscalar $\sigma$ and isovector $\delta$) 
mesons and the baryons, whereas the mass modification of
the charmonium state $\psi(3770)$ is obtained from the medium
change of a dilaton field, which simulates the gluon condensates
of QCD. The anomalous magnetic moments (AMMs) of the baryons
are taken into consideration in the present work.
At finite baryon density, accounting for the Dirac sea effects 
and the AMMs of the baryons, we observe the inverse magnetic catalysis 
(drop in the magnitude of the scalar fields $\sigma$ and
$\zeta$, which are proportional to the strengths of the
light non-strange and strange quark condensates, with increase
in the magnetic field) in magnetized nuclear matter, contrary to the
opposite effect of magnetic catalysis (MC) at zero baryon density.
The inclusion of the hyperons to the nuclear matter 
at finite baryon density is observed
to lead to the effect of magnetic catalysis.
The effect of (inverse) magnetic catalysis is observed to be
enhanced with the increase in the temperature. In the absence
of the Dirac sea effects, the scalar fields ($\sigma$ and $\zeta$)
are observed to remain almost unaffected when the magnetic field
is increased for temperatures T=50 and 100 MeV, whereas, there is
observed to be a rise with $eB$ for T=150 MeV, both for the 
nuclear and hyperonic matter at $\rho_B=\rho_0$. 
The effect due to the isospin asymmetry
is observed to be marginal as compared to the Dirac sea effects.
In the absence of Dirac sea as well as PV contributions,
the neutral open charm meson masses are observed to be 
insensitive to the variation of the magnetic field within 
the chiral model, whereas, the masses of the charged
$D^\pm$ mesons have positive contributions 
from the lowest Landau level (LLL) leading to a monotonic increase
as magnetic field is raised. 
At finite density, the PV mixing effects of the open charm mesons
($D-D^*$ and $\bar D-\bar {D^*}$ mixings) are observed to lead 
to drop in the masses of the pseudoscalar open charm ($D$ and
$\bar D$) mesons. The Dirac sea effects
in nuclear matter for $\rho_B=\rho_0$ are observed to lead 
to an initial increase with increase in the magnetic field, 
followed by a drop of the open charm meson masses 
and the value for which it starts decreasing is smaller with 
increase in the temperature. In hyperonic matter, at $\rho_B = \rho_0$, 
the masses are observed to lead to an increase upto $eB\sim 5 m_\pi^2$, 
upto which the solutions for the scalar fields can be found using 
the weak magnetic field approximation for obtaining the baryon 
self energy incorporating the Dirac sea
contributions by summing over the tadpole diagrams. 

The $\psi(3770)$ mass is observed to show a similar trend 
as for the open charm mesons, when the Dirac effects are 
considered. The mass of the longitudinal component of $\psi(3770)$,
is observed to increase due to the mixing with pseudoscalar
meson, $\eta_c'$. Due to PV mixing, the production cross-section 
of $\psi(3770)$, arising due to the scatterings
of (I) $D^+D^-$ and (II) $D^0 \bar {D^0}$ mesons,
are observed to have distinct peak positions
in the invariant mass spectrum for the higher
value of the magnetic field, $eB=8 m_\pi^2$,
in the magnetized nuclear (hyperonic) matter
for $\rho_B=\rho_0$ as well as for $\rho_B=0$.
This arises due to the difference in the masses 
of the longitudinal and transverse components, 
since the former is modified, 
whereas, the transverse component is unaffected by PV mixing. 
In the presence of DS effects of the baryons, for $eB=8 m_\pi^2$,
it is observed that the production cross-sections of $\psi(3770)$
due to scatterings of (I) $D^+D^-$ and (II) $D^0 \bar {D^0}$ mesons
are peaked at the positions close to the masses of the longitudinal 
and transverse components of $\psi(3770)$ respectively. 
In the absence of the Dirac sea contributions, there are
observed to be additional peaks for the charmonium cross-section
arising from $D^+D^-$ scattering, at the positions close
to the transverse mass of the charmonium state.
For $\rho_B=0$ and $eB=2.5 m_\pi^2$, the peak height 
is observed to be much larger for the $D^0\bar {D^0}$ 
as compared to the value for $D^+D^-$ scattering,
with as well as without DS effects.
The peak heights are observed to drop appreciably as the magnetic 
field is increased.
There is observed to be appreciable increase in the separation 
between the peaks when the temperature is raised to 150 MeV. 
The charmonium decay widths calculated using the field theoretical (FT)
model of composite hadrons are observed to be much larger than the 
values calculated using the $^3P_0$ model, which is reflected
in the production cross-sections of $\psi(3770)$
as an appreciable broadening of the peaks. The present 
study of the charmonium production in isospin asymmetric 
strange hadronic matter at finite temperature and 
in presence of strong magnetic fields
can have observational consequences on the production of the open and
hidden charm mesons in asymmetric ultra-relativistic peripheral 
heavy ion collison experiments.

\end{document}